\documentclass[%
 reprint,%
 amssymb, amsmath,%
aip,cha %
]{revtex4-1}
\usepackage{docs}
\usepackage{dcolumn}
\usepackage{graphicx}
\usepackage{amsfonts}
\usepackage{amsmath}
\usepackage{amssymb}
\usepackage{amsthm}
\usepackage{bm}
\usepackage{xcolor}
\usepackage{import}

\date{\today}

\begin{document}

\title{Mixing of discontinuously deforming media}

\author{L.~D. Smith}
 \email{lachlan.smith@monash.edu}
  \affiliation{ 
Department of Mechanical and Aerospace Engineering, Monash University, Clayton, VIC 3800, Australia
}%
 \affiliation{CSIRO Mineral Resources, Clayton, VIC 3800, Australia}
\author{M. Rudman}%
\affiliation{ 
Department of Mechanical and Aerospace Engineering, Monash University, Clayton, VIC 3800, Australia
}%
\author{D.~R. Lester}
\affiliation{School of Civil, Environmental and Chemical Engineering, RMIT University, Melbourne, VIC 3000, Australia}
\author{G. Metcalfe}
\affiliation{CSIRO Manufacturing, Highett, VIC 3190, Australia}
\affiliation{Department of Mechanical and Product Design Engineering, Swinburne University of Technology, Hawthorn, VIC 3122, Australia}
\affiliation{School of Mathematical Sciences, Monash University, Clayton, VIC 3800, Australia}

\begin{abstract}
Mixing of materials is fundamental to many natural phenomena and engineering applications. The presence of discontinuous deformations -- such as shear banding or wall slip -- creates new mechanisms for mixing and transport beyond those predicted by classical dynamical systems theory. Here we show how a novel mixing mechanism combining stretching with cutting and shuffling yields exponential mixing rates, quantified by a positive Lyapunov exponent, an impossibility for systems with cutting and shuffling alone or bounded systems with stretching alone, and demonstrate it in a fluid flow. While dynamical systems theory provides a framework for understanding mixing in smoothly deforming media, a theory of discontinuous mixing is yet to be fully developed. New methods are needed to systematize, explain and extrapolate measurements on systems with discontinuous deformations. Here we investigate `webs' of Lagrangian discontinuities and show that they provide a template for the overall transport dynamics. Considering slip deformations as the asymptotic limit of increasingly localised smooth shear we also demonstrate exactly how some of the new structures introduced by discontinuous deformations are analogous to structures in smoothly deforming systems.
\end{abstract}


\maketitle

\begin{quotation}
When highly localised discontinuous deformations are added to an otherwise smoothly deforming medium additional topological freedom for particle transport is created, such as `jumping' between streamlines\cite{Christov2010}. We show which structures associated with smooth dynamical systems are preserved in the presence of discontinuous deformation, which are destroyed, and the reason for each. The freedom created by discontinuous deformations enables the creation of new types of transport structures. In particular, we uncover a novel mixing mechanism that can only arise under combined stretching, cutting and shuffling, and demonstrate this in a model fluid flow where the opening and closing of valves combined with a slip boundary condition induces fluid cutting. The mechanism is fundamentally different to both classical smoothly deforming bounded systems and systems with only cutting and shuffling as it exhibits exponential mixing rates in the absence of folding. We introduce the `webs' of Lagrangian discontinuities as a method for studying systems with both continuous and discontinuous deformations that is able to provide a template for the overall transport dynamics, including classical structures found in smoothly deforming systems and new structures that are introduced by discontinuous deformations. We show that these new structures are analogous to structures in smoothly deforming systems by considering a cut as the asymptotic limit of increasingly localised shears.
\end{quotation}

\section{Introduction}

\begin{figure*}
  \begin{minipage}[c]{0.6\textwidth}
    \includegraphics[width=\textwidth]{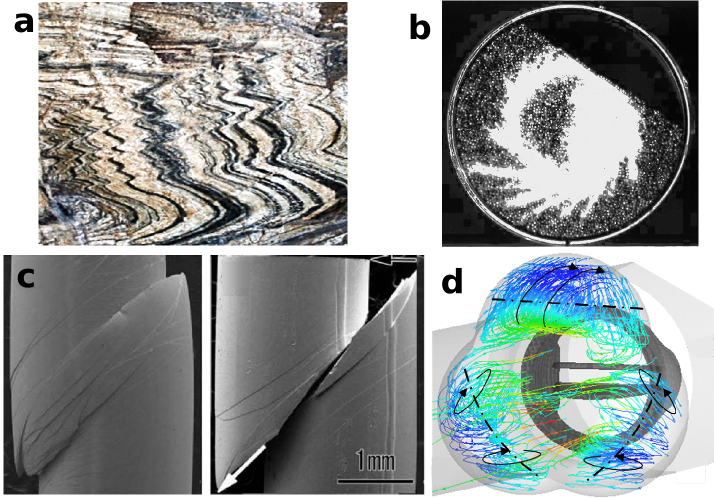}
  \end{minipage}\hfill
  \begin{minipage}[c]{0.35\textwidth}
    \caption{
       Lagrangian discontinuities, which are intrinsic to many materials and valved flows, create the template of mixing and the distribution of mass and energy by cutting and shearing fluid filaments. {\bf a}, Shear banding in geological formations. (Photo Bruce Hobbs) {\bf b}, Shear bands in a granular flow, reproduced with permission from Physica A 233 (1996) \cite{Metcalfe1996}. Copyright 1996 Elsevier. {\bf c}, Shear banding in an alloy, reproduced with permission from Metals 3, 1 (2012) \cite{Louzguine2012}. Copyright 2012 The Multidisciplinary Digital Publishing Institute. {\bf d}, Streamlines through a model heart valve during systole, reproduced with permission from JBSE 6, 2 (2011) \cite{Yagi2011}. Copyright 2011 The Japan Society of Mechanical Engineers.
    } \label{fig:motivation}
  \end{minipage}
\end{figure*}


Dynamical systems theory is the natural language of particle transport and mixing in fluid flows. Since its introduction over three decades ago the signatures of chaos have been found in biological flows \cite{Lopez2001}, geo- and astro-physical flows \cite{Ngan1999,Wiggins2005}, and industrial and microfluidic flows \cite{Nguyen2005}. This approach \cite{Aref} -- termed chaotic advection -- has uncovered the fundamental mechanisms which control fluid mixing and transport in natural and engineered systems.

While chaotic advection largely applies to smoothly deforming materials, there also exist large classes of materials that deform discontinuously, including granular matter, colloidal suspensions, plastics, polymers and alloys \cite{Ottino2000,Olmsted2008}. These materials can exhibit highly localized, discontinuous deformations such as slip surfaces and shear banding (Fig.~\ref{fig:motivation}a-c), which we denote as Lagrangian discontinuities. Understanding the transport and mixing dynamics of these materials is critical to engineering applications such as the development of effective processing methods for granular matter, and understanding natural phenomena such as identifying the deformations which give rise to observed geological formations. 

Moreover, Lagrangian discontinuities may also arise in smoothly deforming materials under certain conditions. For example, in fluid flows with valves and free-slip boundaries, the opening and closing of valves cuts fluid filaments, and the free-slip boundaries allow the cut to be advected into the fluid bulk. Essentially fluid is able to undergo a discontinuous slip deformation analogous to shear-banding. Valves are common to a vast array of applications, including piping networks, vascular networks, multifunctional microfluidic analysis chips, river networks with locks, and the heart (Fig.~\ref{fig:motivation}d). Hence Lagrangian discontinuities arise in a wide range of flows, not just discontinuously deforming materials, and it is important to understand the mixing and transport properties of such flows.

The essence of mixing in materials that deform smoothly is stretching and folding \cite{Aref}, however discontinuous deformation introduces fluid cutting. This seemingly small change has profound consequences for mixing and transport. Cutting introduces an extra degree of topological freedom, hence the cutting action of discontinuities admits new transport and mixing mechanisms, and indeed ergodic mixing is possible in such systems by cutting and re-arranging material elements alone \cite{Juarez2010,Christov2011,Krotter2012,Sturman2012}. While stretching and folding (SF) and cutting and shuffling (CS) can each lead to complete mixing in the sense of the unbounded increase in the interfacial area between marked parts of the continuum, the key difference is that CS does not involve material deformation; the rate of mixing, quantified by the Lyapunov exponent, in SF systems can be exponential, but in CS systems mixing can only be algebraic \cite{Christov2011}. In the terms of ergodic theory CS can only achieve `weak mixing' but not the `strong mixing' characteristic of SF systems. 

Discontinuous deformations cannot be represented as a smooth invertible transformation (diffeomorphism), a building block of classical dynamical systems theory. The stability of periodic points, associated manifolds and interactions captures the global dynamics for smooth systems \cite{Ottino}, however this is not the case for systems with discontinuous deformation due to the added topological freedom. We must therefore extend the scope to include other tools. We introduce the `webs' of Lagrangian discontinuities as a complete tool for studying systems with Lagrangian discontinuities. The `web of preimages' is made up of all points where material will eventually experience a discontinuous deformation and the `web of images' comprises points where the discontinuous deformations are advected into the fluid bulk. They provide additional information to periodic point analysis, creating a template for the overall structure of the system.

In practice, materials that exhibit slip or shear-banding may deform visco-elastically prior to failure, resulting in a mix of CS and SF, and in general a mix of CS and SF arises in all but highly idealized systems. The interactions from even vanishingly weak SF and CS are non-trivial. Mixing in these combined CS and SF systems has been considered \cite{Jones1988,Hertzsch2007,Sturman2012}, but the focus is typically on SF dominated systems where strong mixing occurs. While the coherent structures in CS-only systems can be understood using piecewise-isometries \cite{Sturman2012,Goetz2003}, there has been no previous work on understanding the new structures that can be created in systems with both CS and SF. To illustrate these concepts we consider a model fluid flow in a parameter range where fluid deformation is dominated by CS, but weak stretching can also occur. We show that these combined dynamics produce mixing dynamics that are fundamentally different to those found in SF dominated systems or in CS-only systems. While our model flow exhibits exponential increase in interfacial area normally associated with SF systems, folding is not present. Instead, crossing of invariant curves is achieved via `streamline jumping' produced by the Lagrangian discontinuity, similarly to what occurs in CS-only systems \cite{Christov2010}. Using a simple map comprising composite shear and cutting motions we demonstrate the essential physics which govern mixing of discontinuously deforming media. The generality of this map (consisting of only shearing of material and slip surfaces) means it is universal to the broad class of systems with Lagrangian discontinuities. 

To probe the connection between smooth and discontinuous deformations more deeply, we consider a discontinuous slip surface as the asymptotic limit of increasingly localised shear. The new structures that are introduced by the discontinuous deformation appear as classical coherent structures when the cut is smoothed, but are destroyed in the singular limit towards a cut. These new structures therefore provide the analogues for discontinuous deformations of the classical coherent structures.

\section{A Smooth Flow with Lagrangian Discontinuities}

To illustrate how valves and slip boundaries are able to create discontinuous deformations of fluid that resemble slip surfaces and shear banding of materials we consider a periodically reoriented dipole flow, the Reoriented Potential Mixing (RPM) flow \cite{Lester,Metcalfe2,Trefry}. We choose RPM flow parameters such that CS dominates but some weak fluid stretching is also present. The flow is a 2D incompressible potential (Darcy) flow that approximates flow in porous media, where singularities of the dipole flow mimic valved wellbores used in groundwater applications. The model has been studied both numerically \cite{Lester} and experimentally \cite{Metcalfe2} in the context of chaos and mixing in groundwater flow \cite{Trefry}. This prototpyical model introduces a mix of CS and SF dynamics based on parameters of the flow, and so is well-suited to studying the mixing dynamics from CS-dominated to SF-dominated flows. We highlight the difference between the transport behaviour in the RPM flow compared to classical dynamical systems and CS-only systems.

\begin{figure}[!tb]
\centering
\includegraphics[width=\columnwidth]{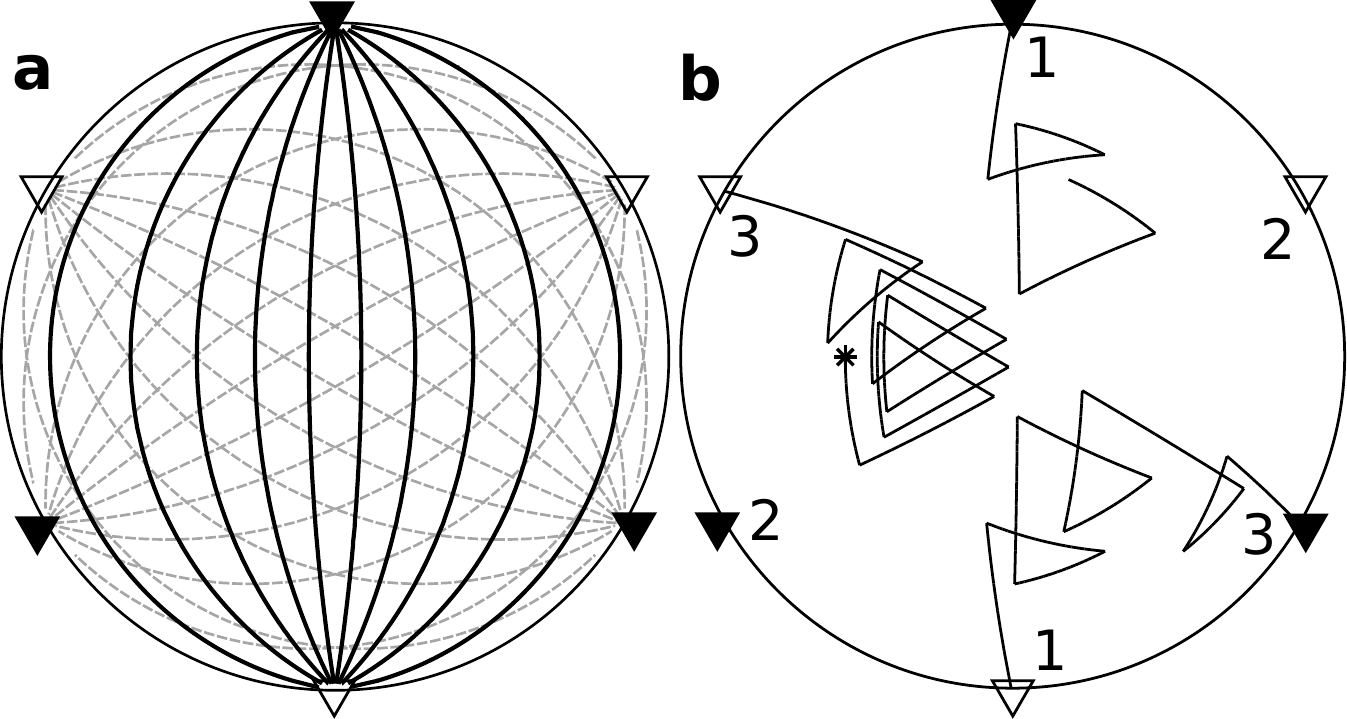}
\caption{The RPM flow. {\bf a}, Streamfunction contours with the dipole (solid/open triangles) in its original (black) and rotated (dashed grey) positions. {\bf b}, Source (solid triangles) and sink (open triangles) positions and a typical particle trajectory in the flow with $(\Theta,\tau)=(2\pi/3,0.1)$, starting at the star.}
\label{fig:RPM_setup}
\end{figure}

The RPM flow is driven by a periodically reoriented dipole flow (Fig.~\ref{fig:RPM_setup}) within the unit circle, the boundary of which is a separating streamline for the unconfined flow, and hence corresponds to a free-slip boundary. After each time period $\tau$ (non-dimensionalized by the emptying time of the domain) the dipole is instantaneously rotated by the angle $\Theta$ to a new position and switched on again. In the limit of vanishing fluid inertia the RPM flow can be considered piecewise steady. Hence the parameters $\tau$, $\Theta$ control the kinematics of mixing and transport in this flow. Previous studies \cite{Lester} have shown the strength of chaos and hence SF increases with increasing $\tau$, and so we expect CS to dominate for small $\tau$, and for large $\tau$ the classical dynamics to dominate. 

The motion of fluid particles within the flow is described by the advection equation 
\begin{equation}
d{\bm{x}}/dt={\bm{v}}({\bm{x}},t)
\label{eq:advection_eq}
\end{equation}
where $\bm{x}$ is the position of the particle and the velocity $\bm{v}$ is controlled by $\Theta$, $\tau$. We denote the position of a particle initially located at the point $\bm{x}$ after time $T$ by $Y_T(\bm{x})$. This advection map is area-preserving since the velocity $\bm{v}$ is incompressible. In this study we consider the RPM flow in closed mode, such that particles are reinjected from the sink to the source along the same streamline. Example trajectories are shown in Fig.~\ref{fig:RPM_setup}b together with the sequence of dipole orientations (see Appendix~\ref{app:particle_tracking} for details on the particle tracking method). 

The steady dipole flow is a 2D incompressible flow, and therefore a one degree-of-freedom Hamiltonian system. While the tools and techniques of Hamiltonian chaos are directly applicable to mixing in 2D incompressible flows, reorientation of the dipole by switching valves combined with the slip boundary condition introduces a Lagrangian discontinuity that is advected into the bulk flow, with consequences for transport that we explore. 

\subsection{CS-dominated transport structures}

\begin{figure*}[p]
\centering
\includegraphics[width=0.9\textwidth]{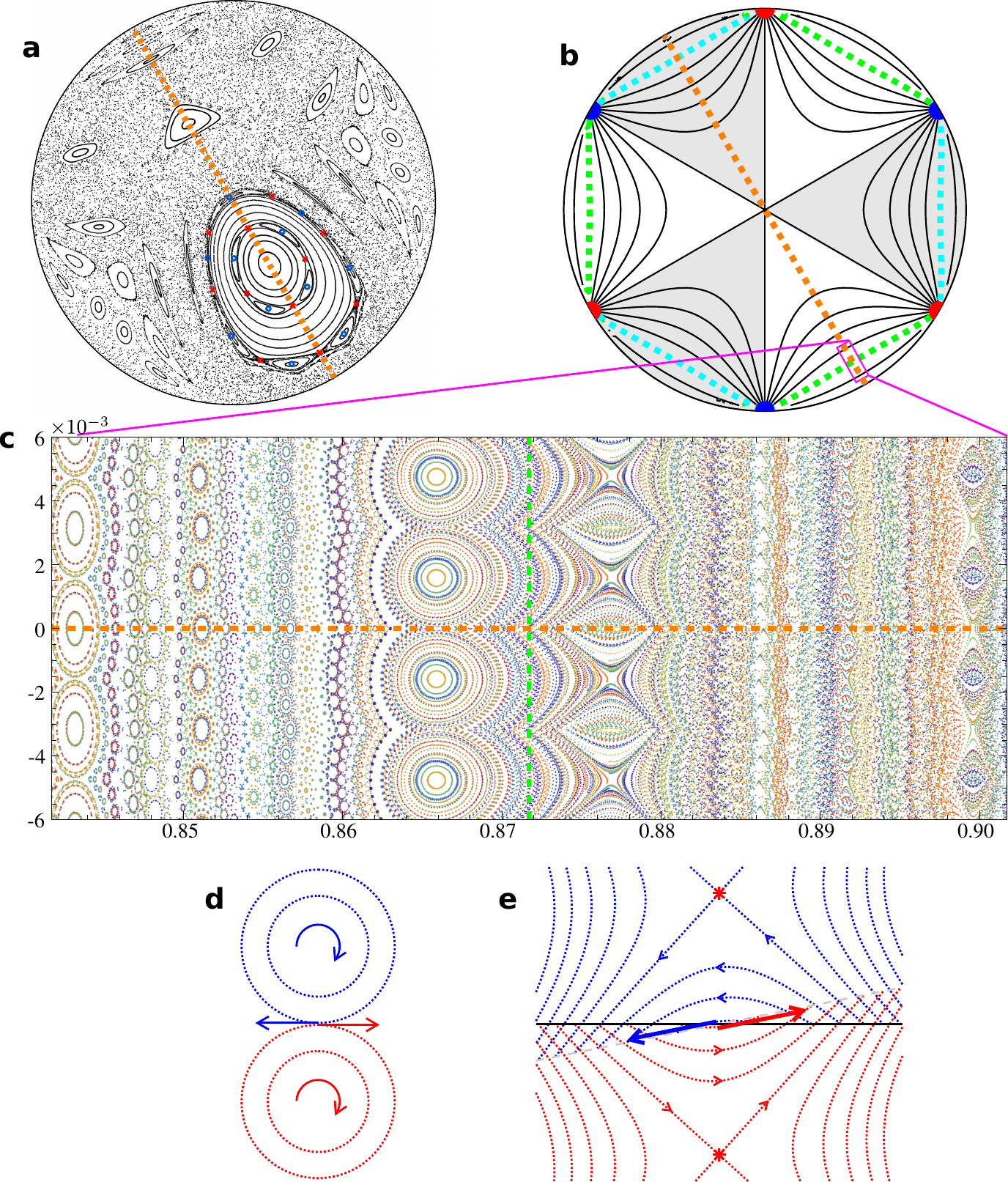}
\caption{Poincar\'{e} sections for the RPM flow. {\bf a}, $(\Theta,\tau)=(2\pi/3,0.2)$, the symmetry line $y=\tan(-\Theta/2)x$ is shown in dashed orange. Chains of elliptic and hyperbolic points are illustrated by blue and red points respectively. {\bf b}, Streamlines of the asymptotic ($\tau \to 0$) flow $\bm{v}_0$ (black), each of which is contained in either the gray or white sectors. The two streamlines with the minimum return time are shown in dashed green and cyan, with the symmetry line again shown in dashed orange. The location of the window used for {\bf c} is shown in pink (not to scale). {\bf c}, A small section of the Poincar\'{e} section for the RPM flow with $(\Theta,\tau)=(2\pi/3,\tau^*)$, with $\tau^*$ defined in the text. The orbit of each particle has a fixed color, showing which structures are part of the same chain. {\bf d,e}, Sketches of the discontinuity produced in the velocity field by (d) co-rotating elliptic islands and (e) hyperbolic points with no elliptic orbit at the centre, such as occurs in the RPM flow.}
\label{fig:RPM_p-sections}
\end{figure*}

To illustrate the difference between CS and SF dominated flows we visualize Lagrangian dynamics via a Poincar\'{e} section; which for temporally periodic flow is a stroboscopic map that captures particle positions over many periods of the flow. Non-mixing regions in the Poincar\'{e} section appear as distinct `islands' (termed KAM-tori) visible in Fig.~\ref{fig:RPM_p-sections}a, whereas mixing regions appear as a topologically distinct `chaotic sea'. Coherent structures for classical Hamiltonian systems are largely organized by the local stability of periodic points, with low-period points playing a dominant role. Elliptic points involve local rotation only and so are locally stable, producing the non-mixing KAM-tori. Conversely, hyperbolic points are locally unstable, with a direction of contraction and a direction of expansion, and generate the chaotic sea. These features are illustrated in the Poincar\'{e} section in Fig.~\ref{fig:RPM_p-sections}a for the RPM flow with $(\Theta,\tau)=(2\pi/3,0.2)$, and are symmetric about the line $y=\tan(-\Theta/2)x$ shown in dashed orange. The Poincar\'{e}--Birkhoff theorem states that in a Hamiltonian system the KAM-tori around an elliptic point will eventually breakup as an alternating chain of elliptic and hyperbolic points, as shown in Fig.~\ref{fig:RPM_p-sections}a by the red (hyperbolic) and blue (elliptic) points. At such large values of $\tau$ the RPM flow is dominated by SF and the organization of structures is essentially the same as a classical Hamiltonian system.

At low values of $\tau$ the Lagrangian discontinuities play a greater role, resulting in structures that are fundamentally different to those encountered in a classical Hamiltonian system. This is seen in Fig.~\ref{fig:RPM_p-sections}c which shows detail of the Poincar\'{e} section for $(\Theta,\tau)=(2\pi/3,5.012\times 10^{-4})$. We primarily focus on this fixed dipole rotation angle $\Theta=2\pi/3$ and fixed switching time $\tau=5.012\times 10^{-4}$, denoted $\tau^*$. The resulting structures violate the Poincar\'{e}--Birkhoff theorem because chains of elliptic points now exist that have no hyperbolic points between them. At such small values of $\tau$ the flow becomes less chaotic over large length scales, and particles closely follow streamlines of the steady flow $\bm{v}_0$ (Fig.~\ref{fig:RPM_p-sections}b) which arises in the limit $\tau\to 0$, consisting of an average of $\bm{v}(\bm{x},t)$ over all dipole orientations. For small values of $\tau$ particle trajectories are perturbed slightly from the streamlines of $\bm{v}_0$, leading to the small-scale structures illustrated in Fig.~\ref{fig:RPM_p-sections}c. This subset is representative of the small-scale behaviour across the entire domain.

\begin{figure*}[!tb]
\centering
\includegraphics[width=0.7\textwidth]{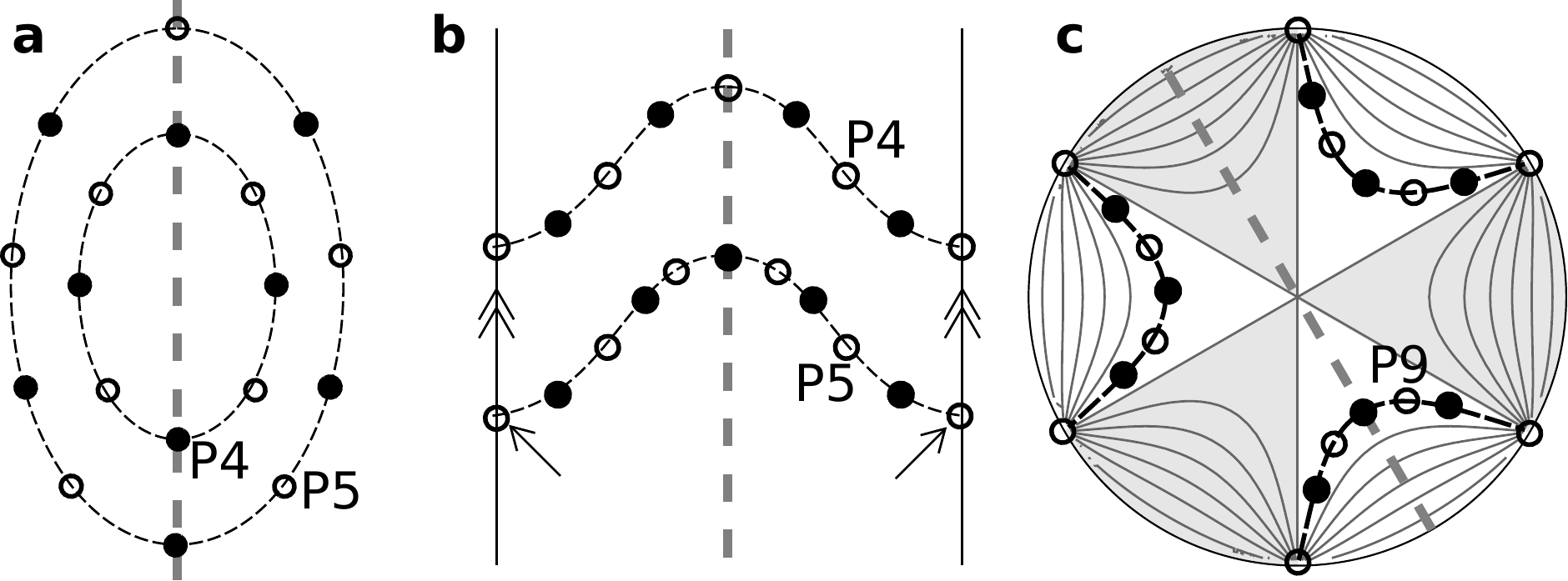}
\caption{Possible arrangements of alternating elliptic and hyperbolic (open/closed circles) period-$N$ points (P$N$) resulting from the breakup of KAM-tori according to the Poincar\'{e}--Birkhoff theorem in systems with a symmetry line (dashed gray). {\bf a},~KAM-tori (dashed black) intersect the symmetry line twice. {\bf b},~KAM-tori intersect symmetry line once and connect to themselves via the periodic boundaries. The points marked by arrows are on the periodic boundary and hence denote the same point. {\bf c},~A sketch of how periodic points could be arranged on a streamline of the flow $\bm{v}_0$ in the limit as $\tau \to 0$ if the Lagrangian discontinuity caused by the dipoles did not exist.}
\label{fig:PB_arrangements}
\end{figure*}

It is convenient to define the streamline return time $t_{\text{return}}$ of the steady flow $\bm{v}_0$ as the time it takes a particle on a given streamline to return to its initial position. The two streamlines with minimum return time are shown as the green and cyan dashed curves in Fig.~\ref{fig:RPM_p-sections}b,c, one in each of the grey and white groups of sectors, and these streamlines separate regions with different Lagrangian structures. On the left side of Fig.~\ref{fig:RPM_p-sections}c there is a fractal distribution of densely packed KAM-tori which form chains that shadow streamlines of $\bm{v}_0$. The periodicity of each of these chains is governed by its resonance with the streamline return time, such that for integers $p,q$, if $\tau/t_{\text{return}}=p/q$ then the period of the island chain is $q$. As the Poincar\'{e}--Birkhoff theorem does not apply to flows with Lagrangian discontinuities, these KAM-tori manifest in the RPM flow as a dense tiling of islands {\em without} hyperbolic points between them. The absence of hyperbolic points is explained via a symmetry argument: as coherent structures must be symmetric about the symmetry line $y=\tan (-\Theta/2)x$ (shown in dashed orange in Fig.~\ref{fig:RPM_p-sections}), any chain of periodic points must also be symmetric about the symmetry line. When the underlying KAM-torus intersects the symmetry line twice, the periodic points must be arranged as in Fig.~\ref{fig:PB_arrangements}a for the odd and even periodicity cases, these configurations are the ones observed in Fig.~\ref{fig:RPM_p-sections}a. It is also possible that the underlying KAM-torus only intersects the symmetry line once, which can be achieved by periodic boundaries as in Fig.~\ref{fig:PB_arrangements}b. This occurs in the RPM flow at low $\tau$ as the KAM-tori are the streamlines of the flow $\bm{v}_0$ (Fig.~\ref{fig:RPM_p-sections}b), forming closed loops connected via the dipoles, each of which only intersect the symmetry line once -- either in the white or gray region. In these cases when the KAM-torus only intersects the symmetry line once, the chains of periodic points must be arranged as in Fig.~\ref{fig:PB_arrangements}b, with one periodic point in each chain occurring on the periodic boundary. For the chains of elliptic points in the RPM flow, if hyperbolic points interleaved them in accordance with the Poincar\'{e}--Birkhoff Theorem, then the dipole positions must be hyperbolic points, as in Fig.~\ref{fig:PB_arrangements}c. However, we will see in Sec.~\ref{sec:valves+wall_slip} that the dipoles are the source of the discontinuous deformations, and any Lagrangian coherent structure that is advected onto the dipole will be destroyed, and hence destroy the entire chain of hyperbolic points. This lack of hyperbolic points allows a denser tiling of islands, a smaller chaotic set, and much slower transport between the islands. With such a dense tiling of islands, large jumps in the velocity are produced at the boundaries of neighbouring islands, as depicted by the arrows representing velocity vectors in Fig.~\ref{fig:RPM_p-sections}d. These jumps are only possible due to the presence of Lagrangian discontinuities.

Similar resonance is observed on the right side of Fig.~\ref{fig:RPM_p-sections}c for chains of hyperbolic points, whose position mirrors the centres of the islands on the left. Like the chains of elliptic points, there is a jump in the velocity field at the point between successive hyperbolic points, demonstrated by the arrows representing opposing velocity vectors at the centre in Fig.~\ref{fig:RPM_p-sections}e. These points represent the centres of leaky regions in which particles are temporarily trapped, i.e. they spend a large amount of time there but eventually leak out and are replaced by particles leaking in from outside. Particles rotate about the centre of the region, as they would on a KAM-torus surrounding an elliptic periodic point, but they are able to move chaotically. In these regions fluid elements experience a mix of stretching and CS, and later we will see that no folding occurs. Strong mixing takes place, an impossibility for both CS-only systems and stretching-only systems.

Later we will see that by approximating the slip deformation by a highly localised smooth shear these periodic point chains can be considered in the limit of smooth systems. As predicted by the Poincar\'{e}--Birkhoff theorem, the point between the elliptic islands in this smoothed approximation is a hyperbolic periodic point, and vice versa. In the limit of a Lagrangian discontinuity these hyperbolic and elliptic points are destroyed by the discontinuity, and hence are denoted as `pseudo-hyperbolic points' and `pseudo-elliptic points' respectively.

\subsection{Valves and wall slip acting as a slip surface}  \label{sec:valves+wall_slip}

\begin{figure*}[p]
\centering
\includegraphics[width=0.75\textwidth]{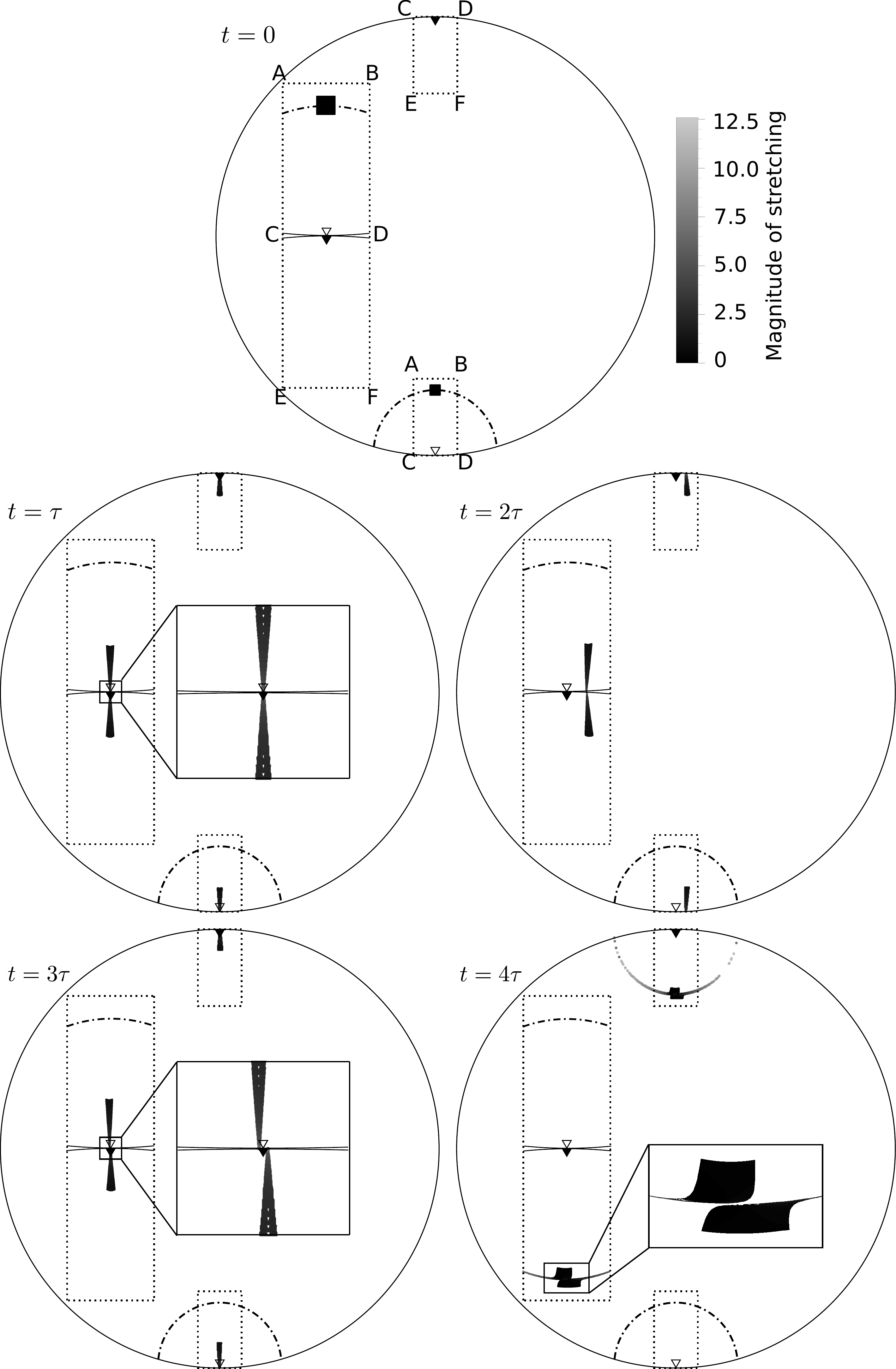}
\caption{The deformation of an initially square set of particles for $(\Theta,\tau)=(2\pi/3,0.02)$ after times $t=n\tau$, $n=0,1,2,3,4$. The domain around the valve, marked by the points A-F, has been placed side by side and inset to make it easier to see the cutting of the fluid filament. Particles are shaded according to the magnitude of local stretching. The position of the Lagrangian discontinuity (where fluid reaches the dipole in one flow period) is shown as the dot-dashed curve. After $t=4\tau$, particles on each side of the Lagrangian discontinuity become separated. (Multimedia view)}
\label{fig:cutting_mechanism}
\end{figure*}

\begin{figure*}[p]
\centering
\includegraphics[width=0.9\textwidth]{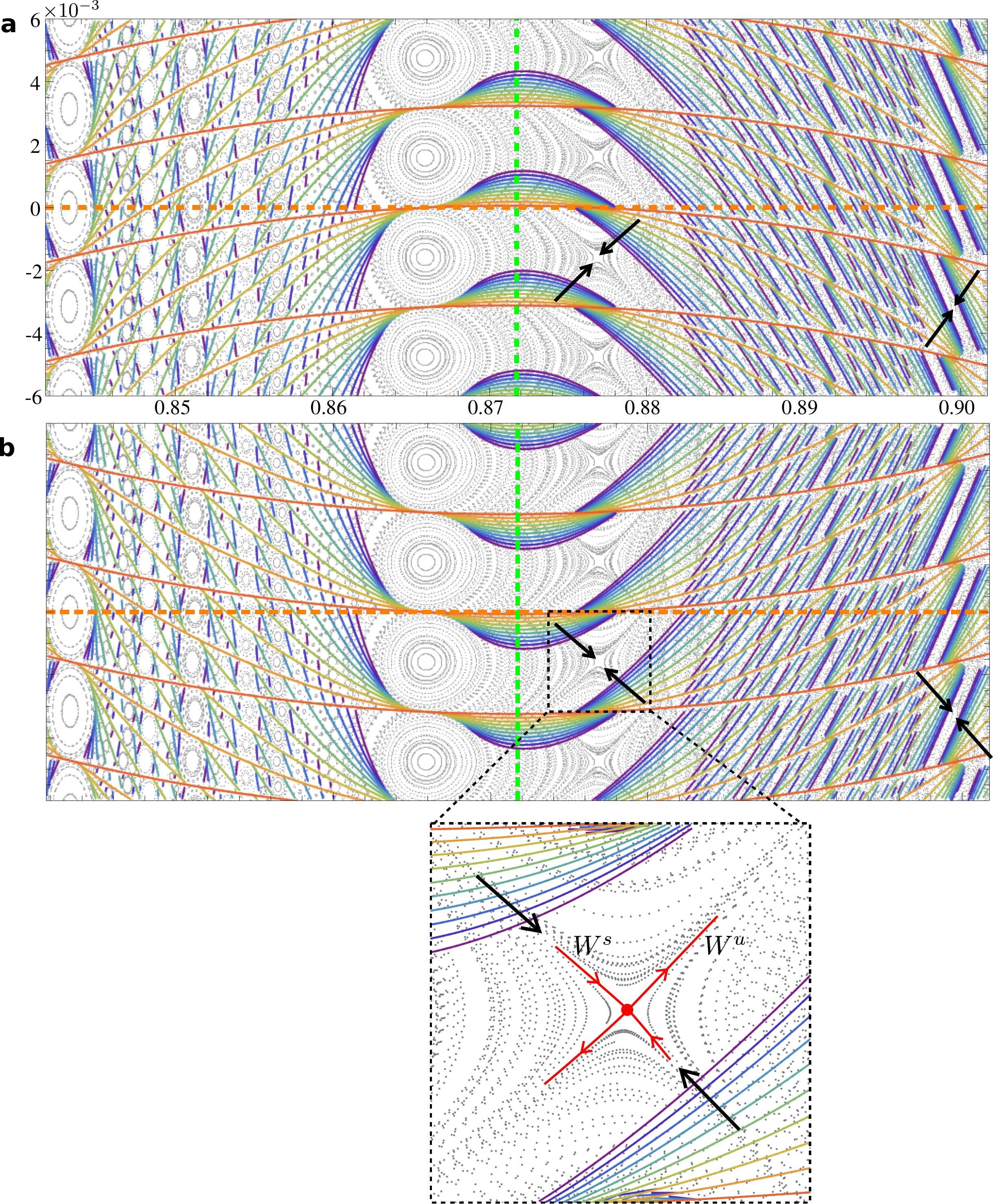}
\caption{The webs of Lagrangian discontinuities $D^s,D^u$ for $(\Theta,\tau)=(2\pi/3,\tau^*)$ shown in the same frame used for Fig.~\ref{fig:RPM_p-sections}c, whose Poincar\'{e} section is shown in faint grey. Curves are coloured from red to purple according to the number of preimages and images respectively. Convergence towards hyperbolic points along the stable and unstable manifolds ($W^{s,u}$) is illustrated by arrows. {\bf a},~Web of preimages. {\bf b},~Web of images, with a blowup around one of the hyperbolic points.}
\label{fig:CL_preimages}
\end{figure*}

Lagrangian discontinuities significantly influence the transport dynamics of incompressible flows. We examine how the opening and closing of valves combined with slip boundary conditions creates Lagrangian discontinuities. Fig.~\ref{fig:cutting_mechanism} shows how this manifests in the RPM flow. Fluid cutting is achieved by closure of the valve, and once closed the slip boundaries allow the disjoint fluid elements to move independently, as occurs via the operation of the reoriented dipoles. If the disjoint fluid does not reconnect to the valve immediately before it is reopened, this discontinuity can be advected through the valve and into the bulk fluid. For flows such as the RPM with non-uniform velocities on the boundary the `cut' appears as a slip deformation. This Lagrangian discontinuity occurs along the curve separating fluid that passes through the valve from the rest of the domain, the dot-dashed curve in Fig.~\ref{fig:cutting_mechanism} that is approximately semi-circular. It can be found by solving $t_{\text{valve}}(\bm{x}) = \tau$, where $t_{\text{valve}}$ is the time it takes a particle to reach the valve. 

Under no-slip boundary conditions, fluid would remain attached to the valve and cutting could not occur. Fluid close to the valve would experience a highly localized shear that can be approximated as a cut when the boundary layer is thin. This idea is discussed in more detail in \S\ref{sec:smooth_cut}

\subsection{The webs of Lagrangian discontinuities}

We now introduce the `webs' of Lagrangian discontinuities, which provide a template for the overall transport dynamics of a system with discontinuous deformations. We demonstrate how the web alone provides information on the nature of the cutting mechanism, the location and stability of periodic points, and pseudo-periodic points.

Lagrangian discontinuities occur at locations where fluid will be cut into disconnected pieces, either via shear banding or slip or via cutting from opening and closing fluid boundaries. In the case of the RPM flow, this occurs at points that are advected onto the valve at the time when the dipole reorients, i.e. the dot-dashed curve in Fig.~\ref{fig:cutting_mechanism}. We can extend this notion by considering locations where fluid will eventually experience discontinuous deformation, not just in the next flow period, but in any subsequent flow period. This corresponds to finding points that will be cut after some number of flow periods, which we achieve by taking successive preimages of the original Lagrangian discontinuity $D_1^s$, the dot-dashed curve in Fig.~\ref{fig:cutting_mechanism}, under the map $Y_\tau$. The result is an infinite {\em web of preimages} 
\begin{equation}
D^s=\{D^s_n=Y_\tau^{-n}(D_1^s), \, n > 0\},
\end{equation}
some of which are shown in Fig.~\ref{fig:CL_preimages}a for the RPM flow. Fluid that straddles the $N$-th preimage $D^s_N$ will remain connected until the $N$-th iteration when it will be cut into disconnected pieces. 

The web of preimages identifies points that will experience discontinuous deformation at some future time. We can also consider points where these discontinuous deformations will propagate throughout the domain, i.e. points where successive iterations under the inverse map $Y_\tau^{-1}$ leads to discontinuous deformation at the dipole. In this case the initial discontinuity consists of points that are advected onto the sink under the {\em inverse} flow, which begins with the dipole in its final orientation (position 3 in Fig.~\ref{fig:RPM_setup}) and polarity reversed. This coincides with the set $D_1^s$ reflected through the symmetry line $y=\tan(-\Theta/2) x$, and we denote it $D_1^u$. Taking successive images of $D_1^u$ under the map $Y_\tau$ yields the {\em web of images} 
\begin{equation}
D^u=\{D^u_n = Y_\tau^n(D_1^u), \, n > 0\},
\end{equation}
some of which are shown in Fig.~\ref{fig:CL_preimages}b. A cutting of fluid that occurs on the $m$-th iteration will appear along the $N$-th image after $N+m$ iterations. The web of images can also be found as the web of preimages of the reverse flow $Y_\tau^{-1}$, and vice versa. Furthermore, due to the RPM flow's reflection-reversal symmetry about the line $y=\tan(-\Theta/2) x$, the web of images is the reflection of the web of preimages through the symmetry line.

\begin{figure}
\includegraphics[width=\columnwidth]{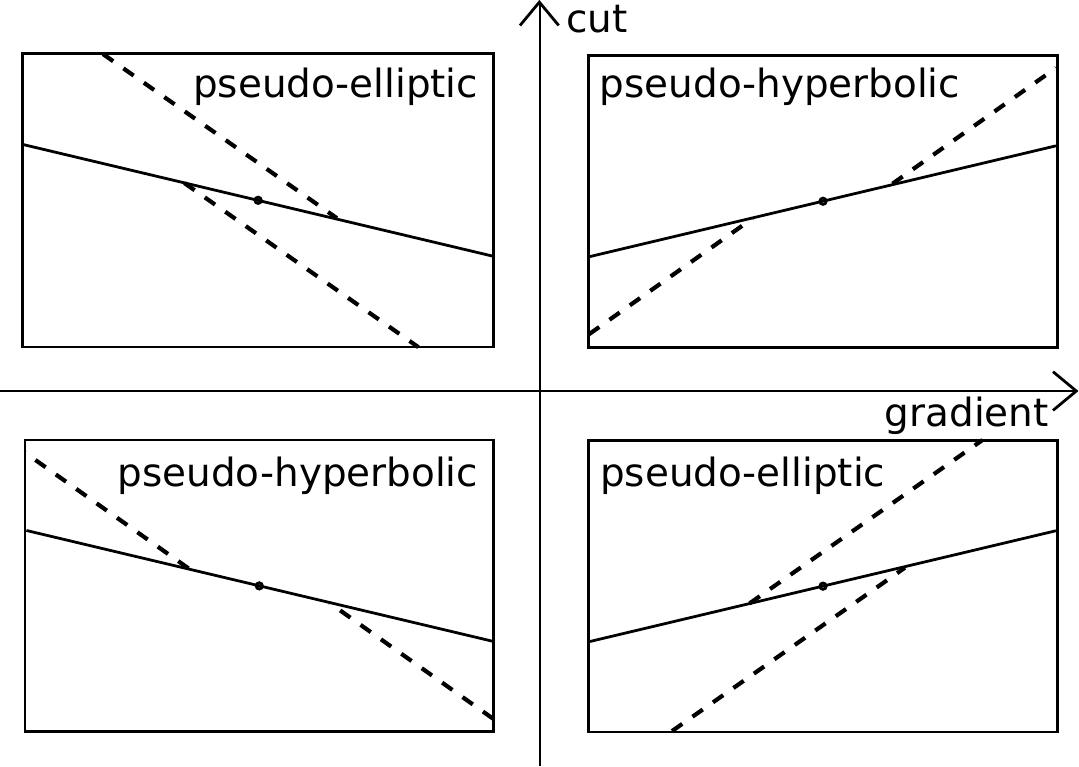}
\caption{Classification of pseudo-periodic points based on the cutting of higher order (dashed) images/preimages by a lower order image/preimage (solid) in the webs of Lagrangian discontinuities. The vertical axis is the signed distance between the end-points of the cut image/preimage, i.e. the displacement of the upper curve relative to the lower curve. Pseudo-hyperbolic points occur where the gradient of the images/preimages has the same sign as the cut, and pseudo-elliptic points occur where the signs are opposite.}
\label{fig:pseudo-per_class}
\end{figure}

Analysing the webs in Fig.~\ref{fig:CL_preimages}a,b, it can be seen that when a higher order preimage (i.e. further away in time, denoted by colours closer to purple) meets a lower order preimage (closer to orange), the higher order curve is cut by the lower order curve. For the web of preimages the upper portion is shifted to the right, and for the web of images it is shifted to the left. This means that cutting of fluid in the flow results in a shift of the upper portion to the left (direction is reversed for the preimages since it is using the inverse flow) where the higher order preimage {\em acts as the slip surface}. This prediction based on the webs can be verified by the inset in the last figure of Fig.~\ref{fig:cutting_mechanism}. Furthermore, it is easy to distinguish classical non-mixing KAM islands in the webs, within which there must exist elliptic periodic points. Hyperbolic periodic points can also be distinguished as points where the webs converge inwards, the web of preimages converges along direction of expansion (unstable manifold $W^u$), whereas the web of images converges along the direction of contraction (stable manifold $W^s$), two examples of this convergence are illustrated by arrows, and this is made clearer by the blown up figure in Fig.~\ref{fig:CL_preimages}b. Elliptic and hyperbolic points could also be found using standard periodic point analysis, however the pseudo-elliptic and pseudo-hyperbolic points cannot be found using conventional methods. These can be found from the webs as points where higher order preimages meet a lower order preimage, and can be distinguished by the way the higher order preimages are cut, shown in Fig.~\ref{fig:pseudo-per_class}. We measure the cut as the signed distance between the two end points, where the sign is given by the difference between the $x$-coordinate of the top and bottom end-points. Pseudo-hyperbolic points occur when the signs of the gradient of the preimages and the cut are the same, and pseudo-elliptic points occur when the signs are different. This same classification can be used for the web of images.

This type of analysis can be applied to general systems with discontinuous deformations, not only fluid flows. The web of Lagrangian discontinuities creates a template for the overall transport dynamics, and provides additional information compared to the standard periodic point analysis.

In Sec~\ref{sec:CSS_map}, we discover the underlying mechanisms that generate the webs and the implications for transport and mixing in general systems with Lagrangian discontinuities.

\section{Basic Mechanisms of Discontinuous Mixing} \label{sec:CSS_map}

\begin{figure*}[!htb]
\centering
\includegraphics[width=\textwidth]{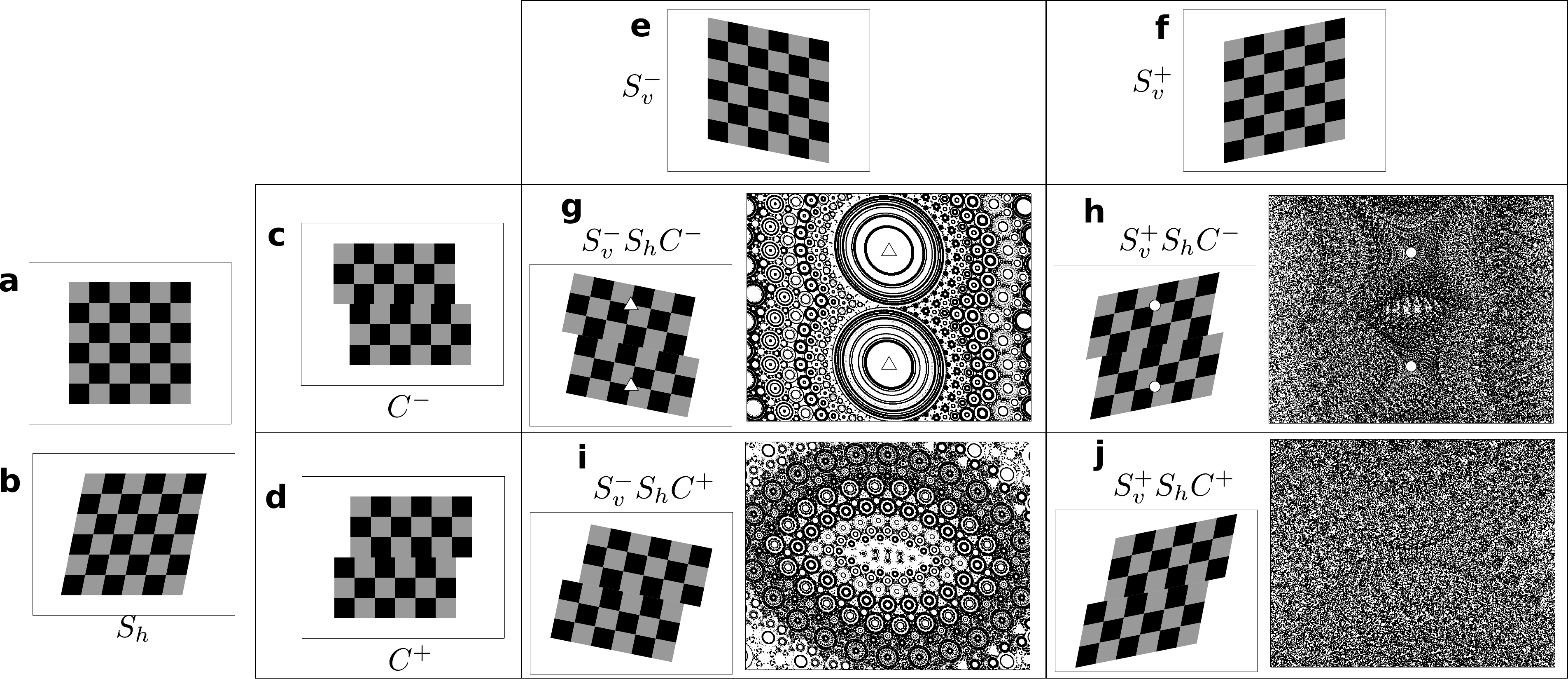}
\caption{The CSS map. {\bf a-f},~The deformations that comprise the $CSS$ map: {\bf a},~no deformation {\bf b},~horizontal shear $S_h$,$\gamma_1=0.2$ {\bf c,d},~positive and negative horizontal cut $C$, $a=\pm0.2$ {\bf e,f},~negative and positive vertical shear $S_v$, $\gamma_2=\mp 0.2$. {\bf g-j},~Combined deformation and associated Poincar\'{e} section for different combinations of the basic deformations. Period-1 points are shown as triangles (elliptic) or circles (hyperbolic), and only exist in (g,h).}
\label{fig:scs_no_rotation}
\end{figure*}

\begin{figure*}[!tb]
\centering
\includegraphics[width=0.9\textwidth]{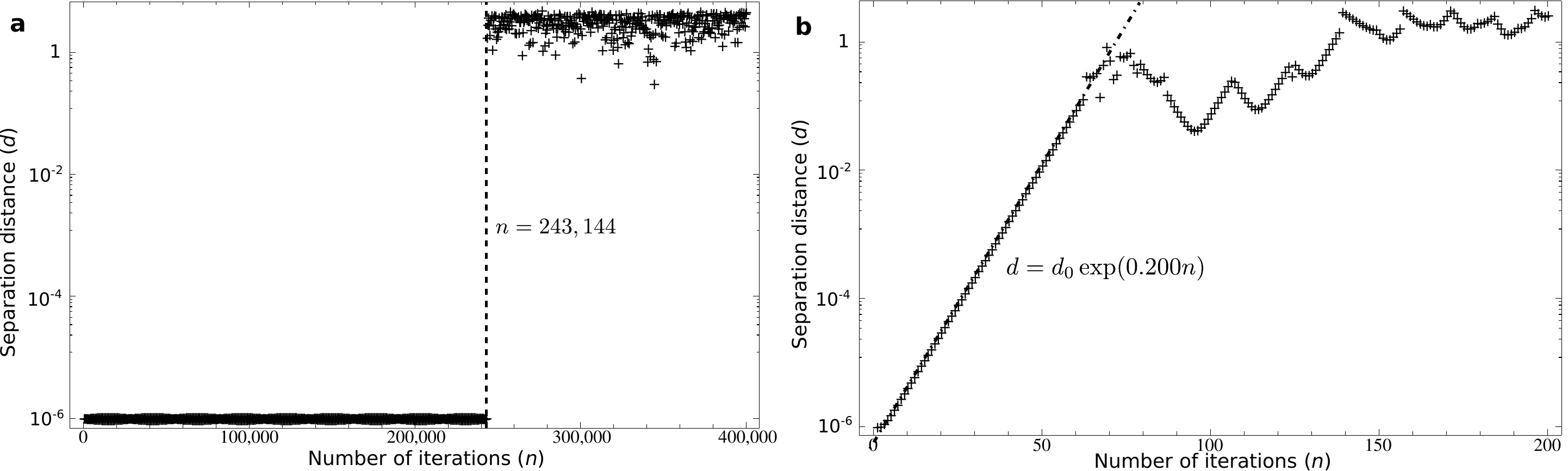}
\caption{The separation of nearby particles in the CSS map. Particles were initially located at $(0,0)$ and $(10^{-6},0)$. {\bf a},~Using the same parameters as in Fig.~\ref{fig:scs_no_rotation}g. The large jump in separation distance is caused by the two particles being cut in opposite directions, which occurs when their $y$-coordinates have opposite signs. {\bf b},~Using the same parameters as in Fig.~\ref{fig:scs_no_rotation}h. The exponent agrees with the Lyapunov exponent associated with the hyperbolic points, which can be found as $\log(\max(\lambda_{1,2})) =  \log \left(\frac{1}{50} \left(51+\sqrt{101}\right)\right) \sim 0.200$ using equation~(\ref{eq:def_tensor_eigen}).}
\label{fig:scs_separation}
\end{figure*}

To elucidate the transport mechanisms associated with cutting and shuffling and stretching and folding motions in materials with Lagrangian discontinuities we consider a simple map which captures the key features present in real systems. In many systems complex mixing dynamics can be idealized as a sequence of shears. If discontinuities are present, then cutting of material can also occur. We introduce a simple map, the Cut-Shear-Shear (CSS) map, such that the net deformation $\Lambda$ consists of a sequence of horizontal cutting (Fig.~\ref{fig:scs_no_rotation}c,d), horizontal shearing (Fig.~\ref{fig:scs_no_rotation}b) and vertical shearing (Fig.~\ref{fig:scs_no_rotation}e,f):
\begin{align}
&C(x,y) = (x+a\,\text{sgn}(y),y),  &S_h(x,y) = (x+\gamma_1y,y), \nonumber \\ 
&S_v(x,y) = (x, y+\gamma_2x),  &\Lambda(x,y) = S_v S_h C (x,y),
\label{eq:CSS_operations}
\end{align}
subject to periodic boundary conditions at $y=\pm 2a/ \gamma_1$. The map $C$ mimics the action of a slip-line as naturally occurs in shear-banding materials, or via a Lagrangian discontinuity which arises from the boundary motion as in the RPM flow. The parameters $a,\gamma_1,\gamma_2$ quantify the magnitude and direction of cutting, horizontal shear, and vertical shear respectively; we will see that varying these parameters can fundamentally change the system. While a large set of simple maps may be constructed from compositions of elementary deformations (cutting, shear, rotation), the CSS map captures the essential features of the dynamics observed in the RPM flow and we will only consider this map here. An exhaustive exploration of these dynamics is required to develop a complete theory of discontinuous mixing, but this is beyond the scope of this study.

For the CSS map there are four topologically distinct combinations for the Lagrangian dynamics. These depend on the signs of $a/\gamma_1,\,a/\gamma_2$ and are shown in Fig.~\ref{fig:scs_no_rotation}g-j. For $a/\gamma_1<0$ the cut and horizontal shear act in opposite directions, and the Poincar\'{e} sections (Fig.~\ref{fig:scs_no_rotation}g,h) reflect the qualitative behaviour of the RPM flow on each side of the minimum return streamline. Conversely, for $a/\gamma_1>0$ (Fig.~\ref{fig:scs_no_rotation}i,j), the cut and horizontal shear act in the same direction. There is therefore no `balancing' of the two deformations and hence no period-1 points. With a positive vertical shear, the behaviour in Fig.~\ref{fig:scs_no_rotation}j at the origin is qualitatively similar to a classical hyperbolic point. The two shears create contraction in one direction and stretching in another. However, the presence of the discontinuities along the $x$-axis and at the periodic boundaries perturbs the motion and `kicks' particles between streamlines, creating widespread chaos. Reversing the vertical shear, the Poincar\'{e} section Fig.~\ref{fig:scs_no_rotation}i has similarities to Fig.~\ref{fig:scs_no_rotation}g. Both have a dense fractal tiling of KAM tori. However there are no period-1 points in this case, and the largest KAM tori are much smaller than the period-1 KAM islands in Fig.~\ref{fig:scs_no_rotation}g. For the remainder of the paper we focus on the cases with $a/\gamma_1<0$ (Fig.~\ref{fig:scs_no_rotation}g,h) to connect the CSS map to the transport and mixing processes of the RPM flow. 

\begin{figure*}
  \begin{minipage}[c]{0.6\textwidth}
    \includegraphics[width=\textwidth]{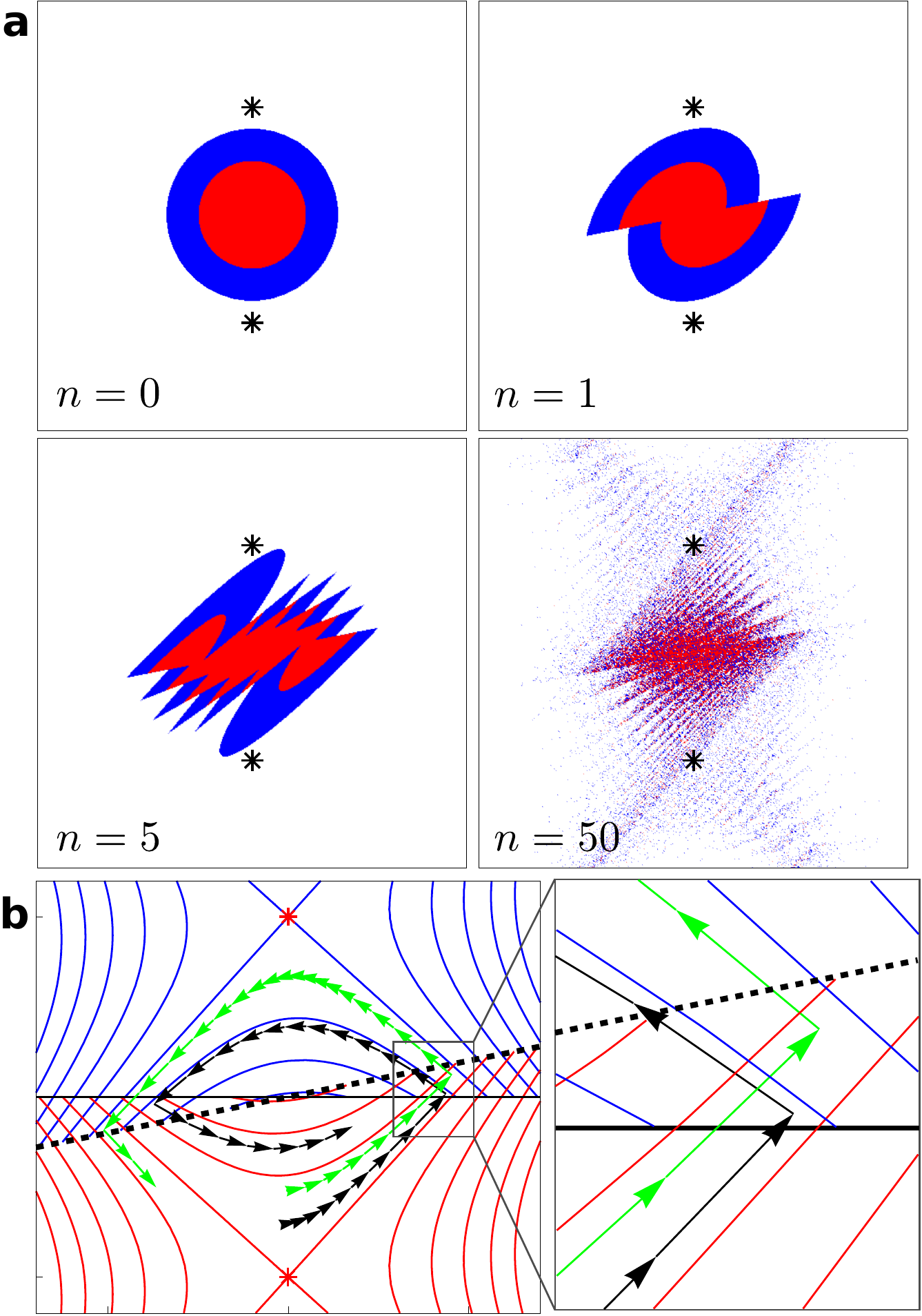}
  \end{minipage}\hfill
  \begin{minipage}[c]{0.35\textwidth}
    \caption{
       Pseudo-elliptic islands in the CSS map with the same parameters as Fig.~\ref{fig:scs_no_rotation}h. The pair of period-1 hyperbolic points are shown as black/red stars. {\bf a}, Dye trace simulation. An initially circular blob of fluid particles (half red, half blue) is iterated under the CSS map $\Lambda$. (Multimedia view). {\bf b}, Invariant curves for the CSS map (red, blue) and particle trajectories (black, green arrows). The dashed black line $y=\gamma_2 x$ defines the region where particles may jump onto new invariant curves.
    } \label{fig:scs_dye_trace}
  \end{minipage}
\end{figure*}


In Fig.~\ref{fig:scs_no_rotation}h the map has two hyperbolic points (circles) which create a pseudo-elliptic island between them, similar to those which occur on the right side of Fig.~\ref{fig:RPM_p-sections}c. To analyse mixing in these regions we performed a dye trace simulation (Fig.~\ref{fig:scs_dye_trace}a), leading to a well-mixed state within 50 iterations of the CSS map. Similar to stretching and folding alone, the combination of CS and stretching leads to exponential separation of nearby particles, as shown in Fig.~\ref{fig:scs_separation}b, however classical Smale `horseshoe' structures associated with fluid folding do not arise. Rather, cutting and rearranging plays the role of folding. This simple cutting, shuffling and stretching process is responsible for fluid mixing in the region of the RPM flow contained between the minimum return streamline and the circular boundary (Fig.~\ref{fig:RPM_p-sections}b). While this behaviour has not been recognized in previous studies, it is likely to arise in most systems with combined stretching and CS. 

To elucidate the mechanisms which drive chaotic motion and leaking of particles, we consider the curves
\begin{equation}
\frac{x^2}{\gamma_1} - \frac{(y\pm \frac{a}{\gamma_1})^2}{\gamma_2} + x\left(y\pm \frac{a}{\gamma_1}\right) = c, \quad c \in \mathbb{R}
\end{equation}
that are invariant under the CSS map, shown as blue and red in Fig.~\ref{fig:scs_dye_trace}b. The system can be thought of as two affine linear systems `glued' together along the $x$-axis, such that the composite CSS map can be expressed as
\begin{equation}
\Lambda(x,y) = \left( \begin{matrix}
1 & \gamma_1 \\
\gamma_2 & 1+\gamma_1\gamma_2
\end{matrix} \right)
\left(\begin{matrix}
x\\
y
\end{matrix} \right) 
+a\, \text{sgn}(y)\left( \begin{matrix}
1 \\
\gamma_2 
\end{matrix} \right).
\end{equation}
A particle above the $x$-axis will follow the blue invariant curves, and below it will follow the red. The key is that when crossing the $x$-axis, e.g. from red to blue, it will not follow the blue invariant curve until the start of the next iteration, allowing it to jump onto a range of blue curves (any of those before the line $y=\gamma_2 x$ which is shown in dashed black). This `streamline jumping' is directly due to the discontinuity along the $x$-axis, and causes the widespread chaos. Successive jumps outward can result in particles escaping the pseudo-elliptic island, and likewise particles from outside are able to jump in.

\begin{figure*}[!htb]
\centering
\includegraphics[width=0.9\textwidth]{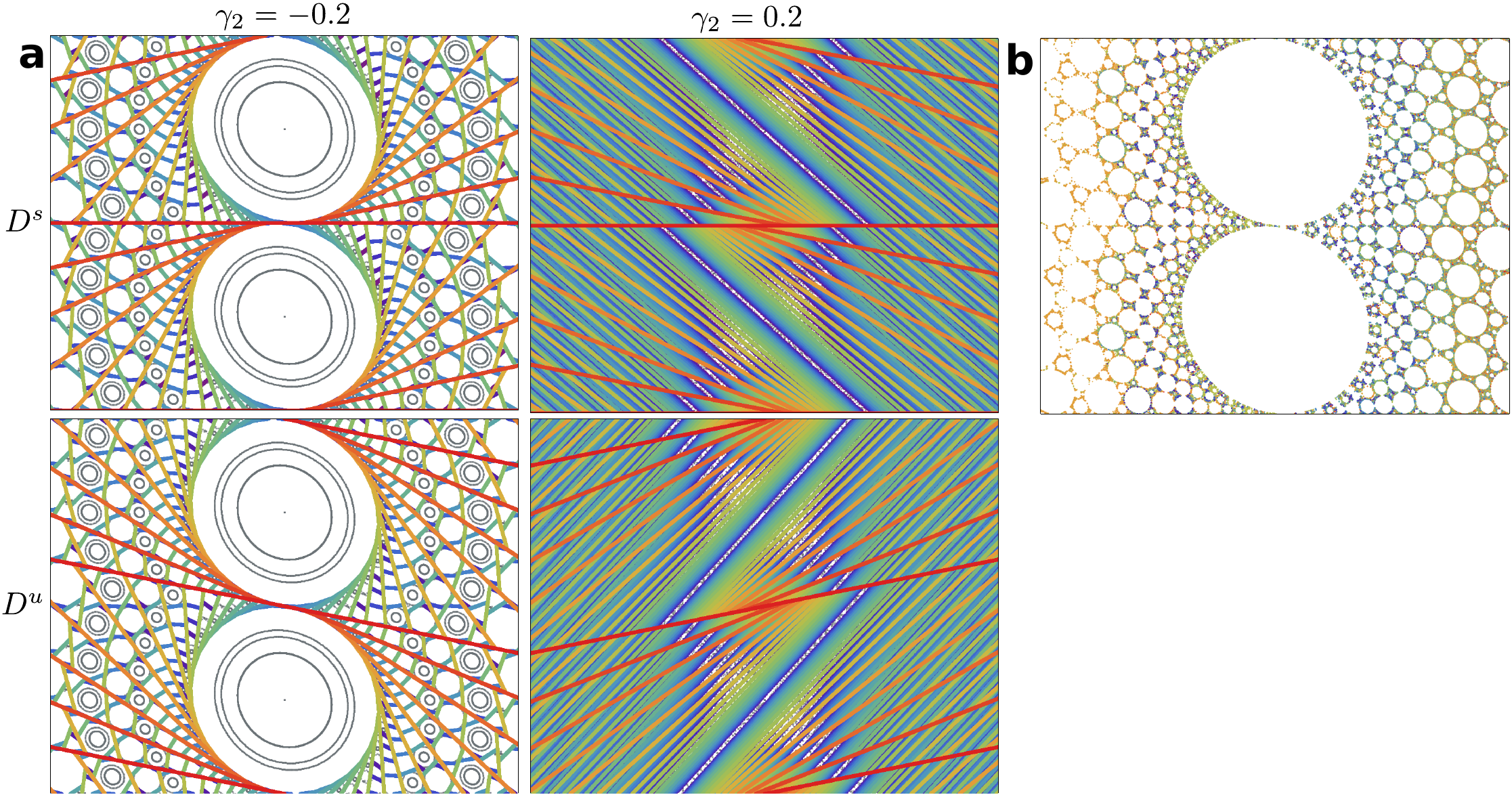}
\caption{Webs of Lagrangian discontinuities for the CSS map. {\bf a}, The webs of preimages, $D^s$, and images, $D^u$, of Lagrangian discontinuities, colored according to number of pre/post-images (red-purple) with the same parameters as Fig.~\ref{fig:scs_no_rotation}g,h. The corresponding Poincar\'{e} sections from are shown in grey. {\bf b}, Tracking a single particle initially located at the origin for 2,000,000 iterations (colored from purple to red) of the CSS map with the same parameters as Fig.~\ref{fig:scs_no_rotation}g shows that the webs of Lagrangian discontinuities coincide with the chaotic set between the non-mixing islands.}
\label{fig:scs_web}
\end{figure*}

Fig.~\ref{fig:scs_no_rotation}g shows that by simply reversing the direction of the vertical shear, the two period-1 points become elliptic (triangles) and a dense fractal tiling of non-mixing islands is created. This is qualitatively similar to the dynamics observed in the RPM flow on the left side of Fig.~\ref{fig:RPM_p-sections}c. While there is no mixing within the islands, Fig.~\ref{fig:scs_web}b shows that a particle initially located at the origin will eventually trace out the entire region between them, forming the chaotic set. In this case the combination of the two shears results in a rotation and hence the distance between initially close particles will not grow until the cut separates them, as shown in Fig.~\ref{fig:scs_separation}a. This case is therefore a CS-only system, consisting of only elliptic rotation and cutting, and hence is associated to a piecewise isometry. The complex structures for the webs of the CSS map are anticipated by the theory of piecewise isometries \cite{Goetz2003,Sturman2012}. They are multifractal and can be very difficult to resolve numerically at small length scales. The structure of the web depends only on the rotation angle generated by the elliptic points. Similar webs appear in other contexts, including the outer billiards map \cite{Moser1978}, overflow in digital filters \cite{Davies1995}, and kicked Hamiltonians \cite{Ashwin1997}, all of which are driven by elliptic rotations and discontinuous deformation, much the same as the CSS map in this case.

Repeating the process of finding successive preimages of Lagrangian discontinuities resolves the webs of Lagrangian discontinuities (Fig.~\ref{fig:scs_web}a). As in the RPM flow, the webs form a template for the system, revealing the nature of the cutting mechanism, the periodic points, and the pseudo-periodic points. It can also be seen that the web Fig.~\ref{fig:scs_web}a coincides with the chaotic set Fig.~\ref{fig:scs_web}b.

The dynamics of the CSS map clearly demonstrates how Lagrangian discontinuities create two new types of transport mechanism that are not possible in classical SF systems. The first is via pseudo-elliptic islands, where the Lagrangian discontinuity enables jumping between invariant curves and mixing in a leaky region via a process of stretching, cutting and shuffling. Exponential separation of nearby particles, and hence strong mixing, is achieved due to the presence of stretching. This leads to more rapid mixing than can be achieved in a CS-only system. The second mechanism is in the measure-zero set amongst the densely packed elliptic regions given by the webs of preimages and images. In this set particles travel ergodically, experiencing a combination of rotation and cutting deformations. Such a dense tiling of elliptic points without any hyperbolic points is impossible in classical Hamiltonian systems, but is characteristic of other CS-only systems (piecewise isometries) that are driven by rotation and cutting. We have demonstrated that these transport behaviours can occur in real systems even when the base system is conservative, and can play a fundamental role in the overall dynamics of transport and mixing.

\section{The Impact of Non-linear Deformations}

\subsection{The non-linear CSS map}

\begin{figure*}[!htb]
\centering
\includegraphics[width=0.9\textwidth]{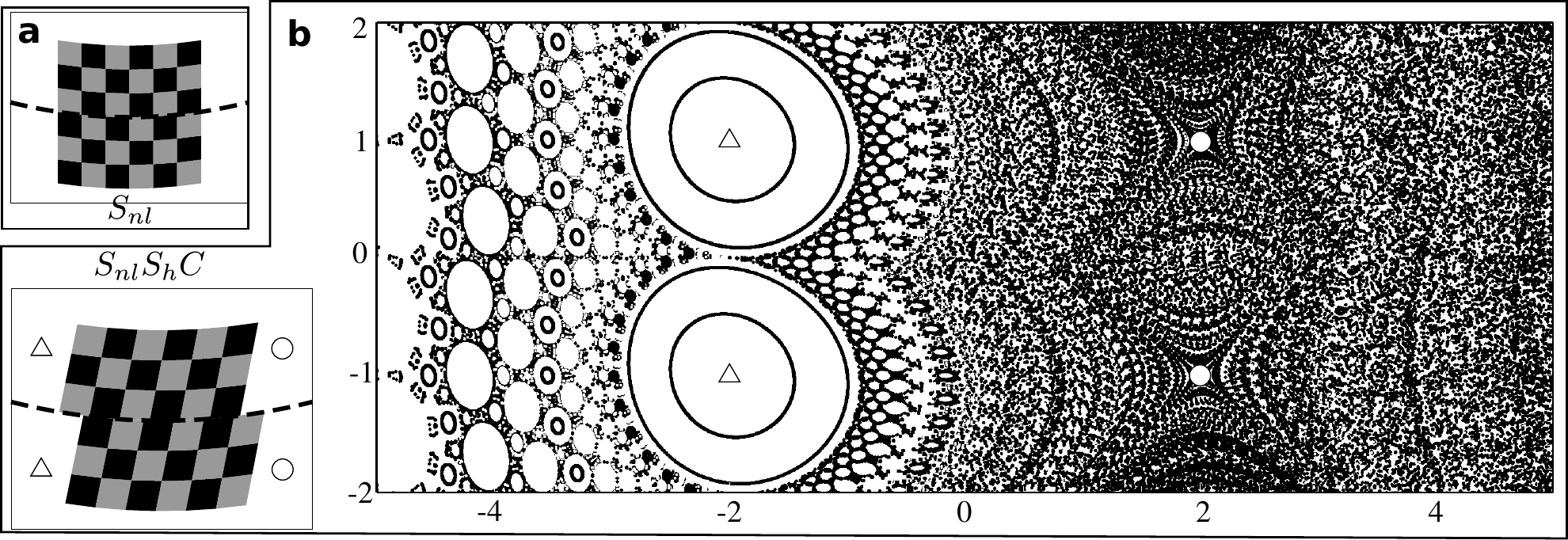}
\caption{The non-linear CSS map equation~(\ref{eq:CSS2_operations}). The initial condition, cut and horizontal shear are identical to those in Fig.~\ref{fig:scs_no_rotation}a-c. {\bf a},~Non-linear vertical shear $S_{nl}$, with $f(x)=0.05x^2 - 0.2$ (dashed). {\bf b},~The combined deformation and associated Poincar\'{e} section. Period-1 points are shown as triangles (elliptic) or circles (hyperbolic). Note the similarity between this Poincar\'{e} section and Fig.~\ref{fig:RPM_p-sections}c.}
\label{fig:scs_model_2}
\end{figure*}

The CSS map only considers linear shear, but in most applications stress is non-uniform, resulting in varying shear rates even in Newtonian materials, with the potential for much stronger variation if yield stress or shear rate dependence is present. In fluids they arise in flows that have non-uniform velocity profiles, which occurs in the RPM flow based on the non-linear return time distribution of the asymptotic flow $\bm{v}_0$. We extend the CSS map by replacing the linear vertical shear $S_v$ with the non-linear vertical shear $S_{nl}$ (Fig.~\ref{fig:scs_model_2}a):
\begin{equation}
S_{nl}(x,y) = (x,y+f(x)), \quad \Lambda_2(x,y) = S_{nl} S_h C(x,y).
\label{eq:CSS2_operations}
\end{equation}
In Fig.~\ref{fig:scs_model_2} $f(x)=0.05x^2 - 0.2$ (dashed) is quadratic such that the shear is negative for $x<0$ and positive for $x>0$. Locally the non-linear CSS map behaves similarly to the linear CSS map, but includes regions of positive and negative vertical shear. We do not show it here, but by approximating the non-linear function $f(x)$ with a piecewise linear function, the linear CSS map can make a piecewise approximation to the non-linear CSS map to arbitrary accuracy. Thus the Poincar\'{e} section for the non-linear map (Fig.~\ref{fig:scs_model_2}b) has both the novel pseudo-elliptic islands and the dense tiling of non-mixing islands from Fig.~\ref{fig:scs_no_rotation}g,h. Using this quadratic function $f(x)$, the Poincar\'{e} section for the non-linear model (Fig.~\ref{fig:scs_model_2}b) captures the key features of the RPM flow (Fig.~\ref{fig:RPM_p-sections}c). We are therefore able to understand the mechanisms driving the transport in the RPM flow at a fundamental level, and because any function $f(x)$ can be approximated by sequences of the up- or down-going vertical shears, these basic transport structures will appear for every non-linear shear profile.

\subsection{Segregated Mixing}

\begin{figure*}
  \begin{minipage}[c]{0.6\textwidth}
    \includegraphics[width=\textwidth]{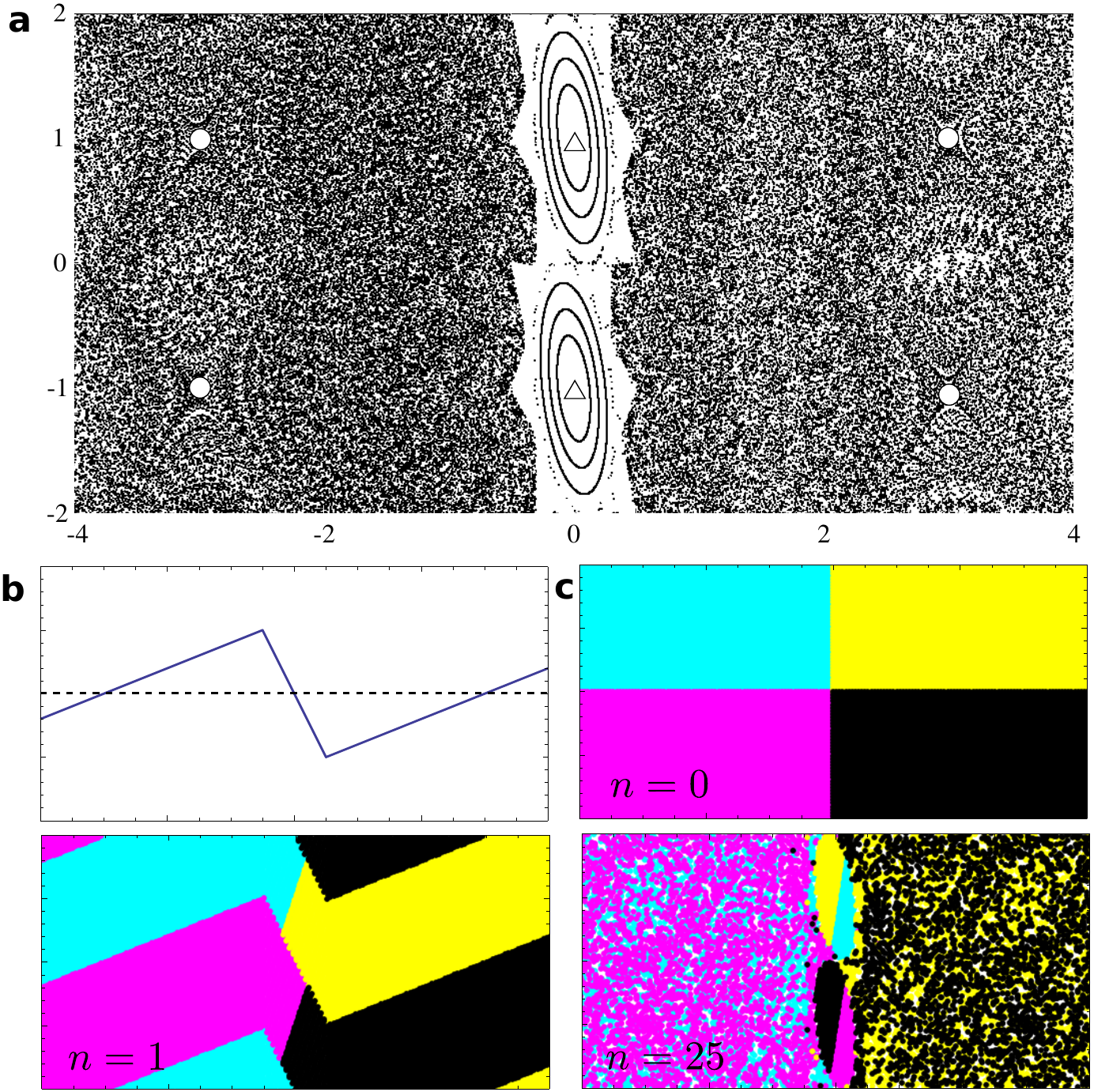}
  \end{minipage}\hfill
  \begin{minipage}[c]{0.35\textwidth}
    \caption{
       Segregation of mixing regions using a piecewise-linear vertical shear. {\bf a},~The Poincar\'{e} section generated using the piecewise linear function shown in {\bf b} as $f(x)$ in equation~(\ref{eq:CSS2_operations}). Hyperbolic and elliptic period-1 points are shown as circles and triangles respectively. {\bf c},~A dye trace simulation with four initially separated colours of dye. Dye is shown after 0,1,25 iterations respectively. (Multimedia view).
    } \label{fig:scs_model_3}
  \end{minipage}
\end{figure*}


The quadratic CSS map shows that it is possible to combine the mixing capabilities of pseudo-elliptic points with non-mixing regions using non-linear vertical shears. By using more complex functions $f(x)$ in equation~(\ref{eq:CSS2_operations}) it is possible to engineer systems that have multiple mixing and non-mixing regions. Pseudo-elliptic islands are created anywhere that the function $f(x)$ has $x$-intercepts with a positive gradient, and dense non-mixing islands are created when the $x$-intercepts have negative gradient. It is therefore simple to design systems that combine mixing and non-mixing regions by controlling the location and gradients of $x$-intercepts of $f$. An example is shown in Fig.~\ref{fig:scs_model_3}, where segregated mixing is achieved using the piecewise linear function $f(x)$ shown in Fig.~\ref{fig:scs_model_3}b. Mixing regions are created on the left and right via pseudo-elliptic points, and in the center there is a thin non-mixing region. The fact that the non-mixing islands form a dense tiling means that they form vertical barriers to transport with very slow leakage across them, separating the two mixing regions. The segregated mixing is clearly illustrated by the dye trace in Fig.~\ref{fig:scs_model_3}c, within $25$ iterations the left and right sides are well mixed, but there is little intermixing between the left and the right.

\section{Transition from SF to CS} \label{sec:smooth_cut}

\begin{figure}[!tb]
\centering
\includegraphics[width=0.9\columnwidth]{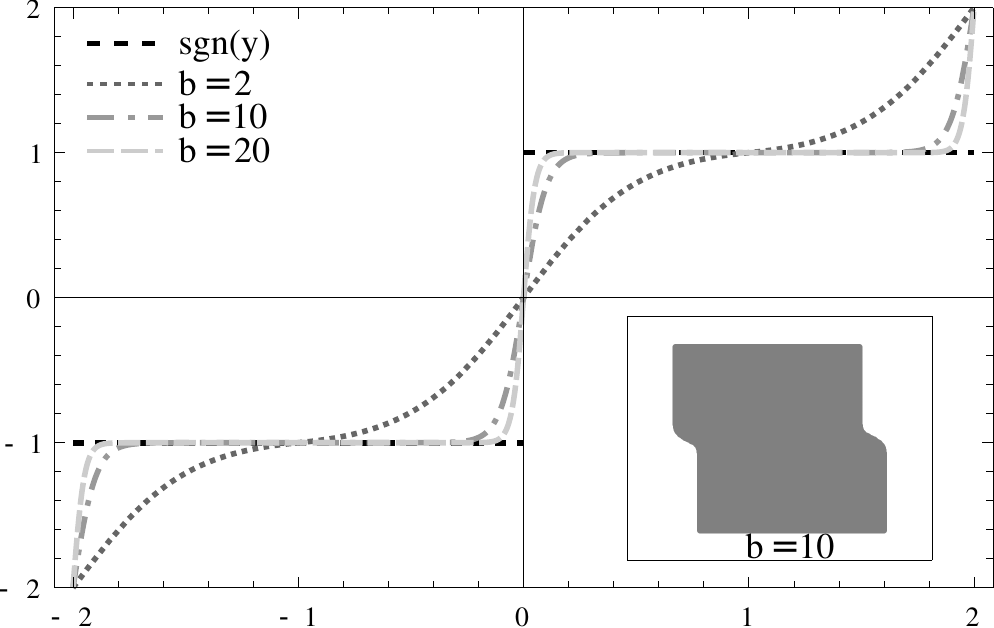}
\caption{The smooth functions $g_b(y)$ used to approximate the discontinuous function $\text{sgn}(y)$. Inset is the deformation of a square by the smoothed cut $\mathcal{C}_b$ with $b=10$.}
\label{fig:smooth_cut_fn}
\end{figure}

\begin{figure*}
  \begin{minipage}[c]{0.65\textwidth}
    \includegraphics[width=\textwidth]{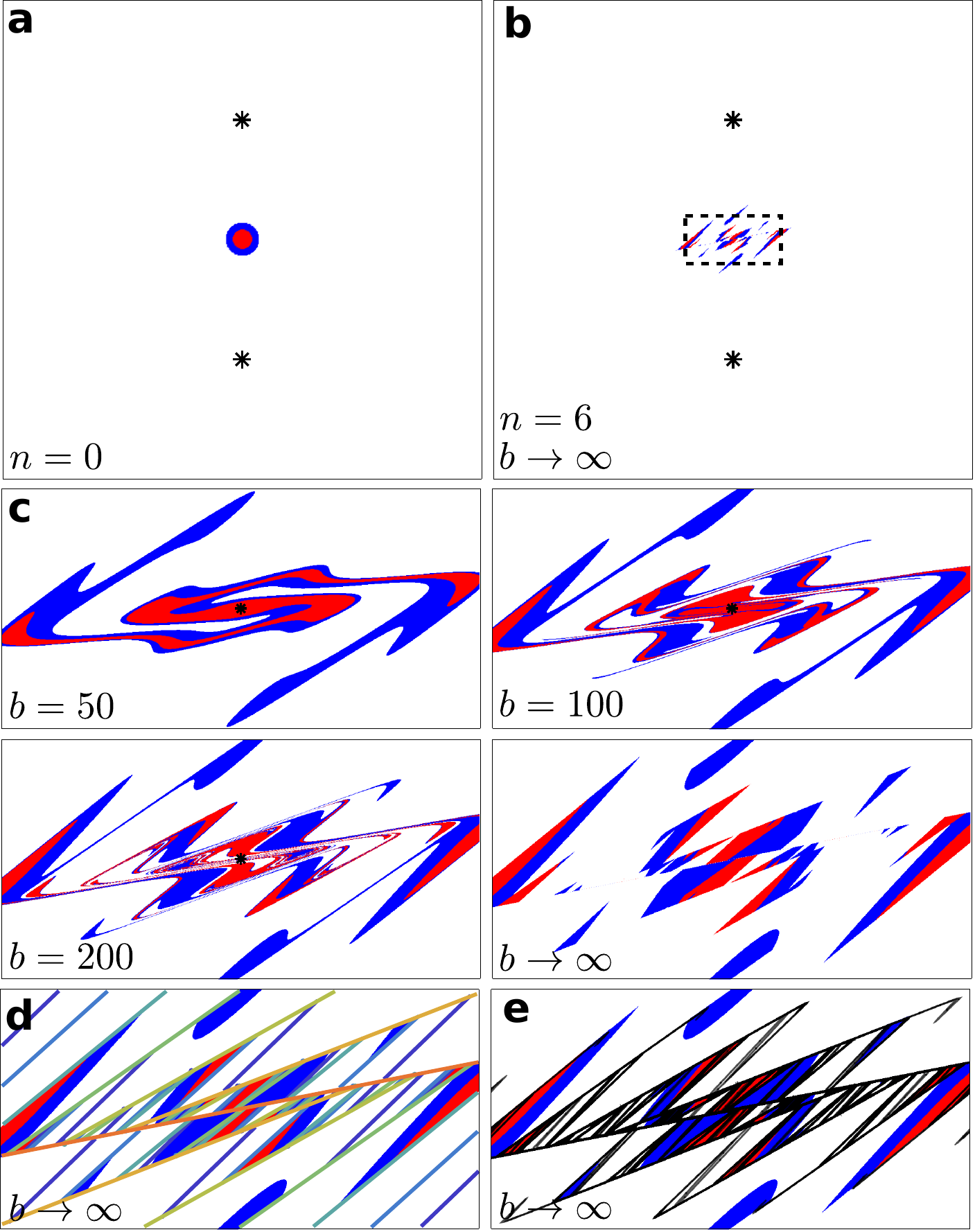}
  \end{minipage}\hfill
  \begin{minipage}[c]{0.3\textwidth}
    \caption{
       The effect of the smoothing parameter $b$ on a dye trace simulation with the same parameters as Fig.~\ref{fig:scs_no_rotation}h. {\bf a}~The initial configuration of the dye is a red disk enclosed within a blue annulus. The hyperbolic period-1 points at $(0,\pm 1)$ are shown as black stars. {\bf b}~The location of the dye after 6 iterations of the CSS map $\Lambda$. The dashed rectangle is the domain used for c, d and e. {\bf c}~Increasing the smoothing parameter $b$, converging towards the discontinuous limit. {\bf d}~The dye trace for the CSS map with the web of images of the Lagrangian discontinuity, $D^u$. Further from red is more distant in time. This web coincides with the location of the striations at high values of $b$. {\bf e}~The dye trace for the CSS map shown with the unstable manifold (black) associated with the hyperbolic point at the origin for the smoothed map $\Lambda_b$ with $b=2000$. As $b$ approaches infinity the unstable manifold converges to the same web of post-images of the Lagrangian discontinuity in d. Likewise the stable manifold converges to the web of pre-images of the Lagrangian discontinuity.
    } \label{fig:smooth_dye_trace_origin}
  \end{minipage}
\end{figure*}


\begin{figure*}[!tb]
\centering
\includegraphics[width=0.95\textwidth]{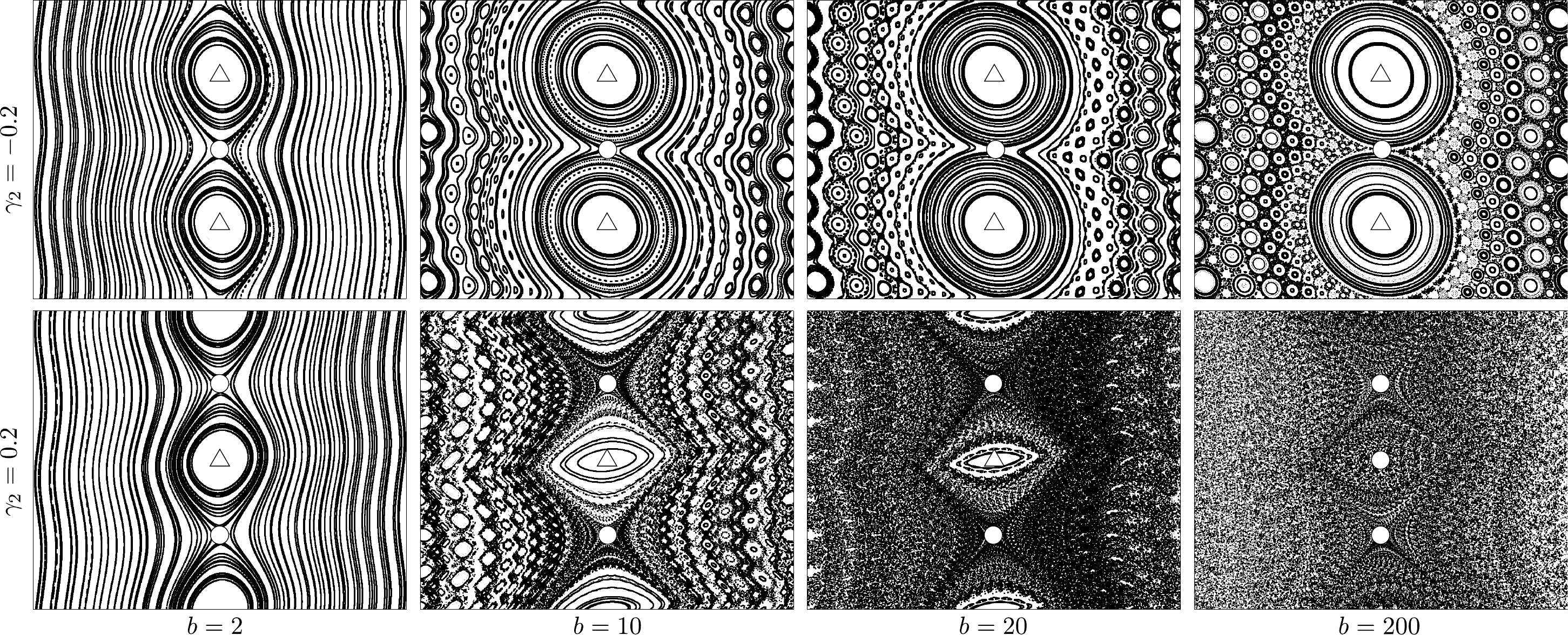}
\caption{Convergence of the Poincar\'{e} sections for the smoothed map $\Lambda_b$ toward the corresponding Poincar\'{e} sections Fig.~\ref{fig:scs_no_rotation}g,h of the discontinuous map $\Lambda$. Hyperbolic and elliptic period-1 points are shown as circles and triangles respectively, and are located at $(0,\pm 1)$ and $(0,0)$. For $\gamma_2=0.2$ (bottom row), at $b\approx 101$ the periodic point at the origin bifurcates from elliptic to hyperbolic (see Appendix~\ref{app:ppoints_smooth} for more details).}
\label{fig:smooth_cut}
\end{figure*}

\begin{figure*}[!tb]
\centering
\includegraphics[width=0.9\textwidth]{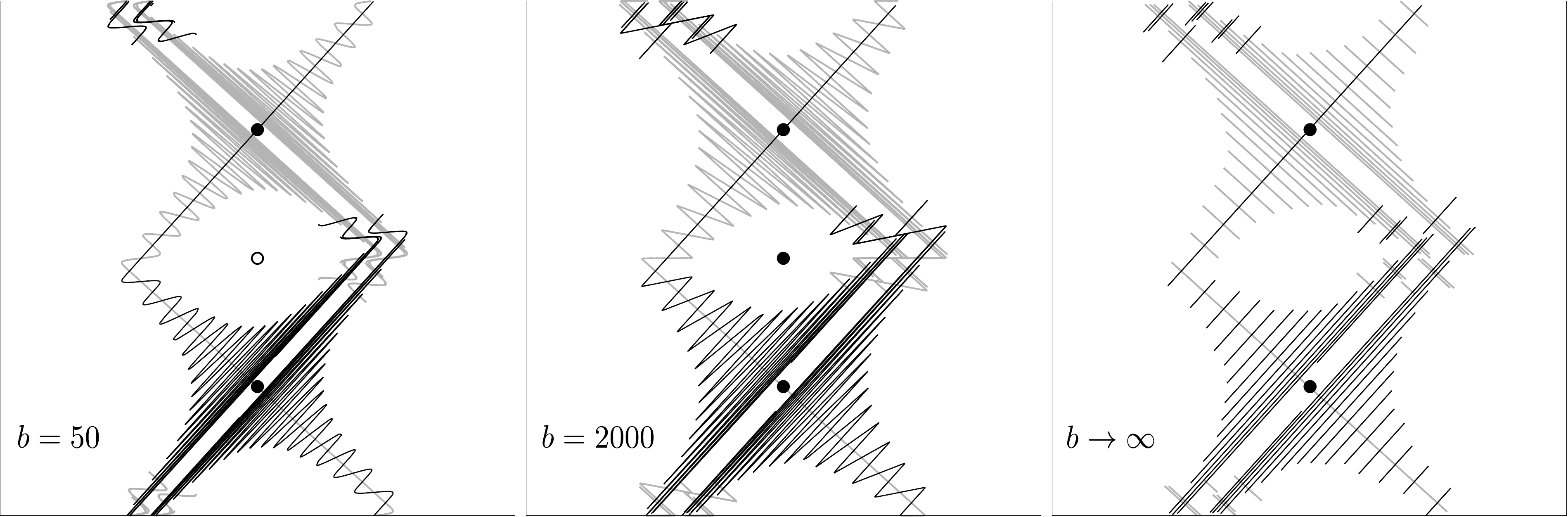}
\caption{The stable (gray) and unstable (black) manifolds associated with the hyperbolic points (solid circles) located at $(0,-1)$ and $(0,1)$ respectively, with the same parameters as Fig.~\ref{fig:scs_no_rotation}h. The first two plots are for the smoothed map $\Lambda_b$ ($b=50,2000$), and the final plot is for the CSS map $\Lambda$.}
\label{fig:smooth_manifolds}
\end{figure*}

\begin{figure*}
  \begin{minipage}[c]{0.6\textwidth}
    \includegraphics[width=\textwidth]{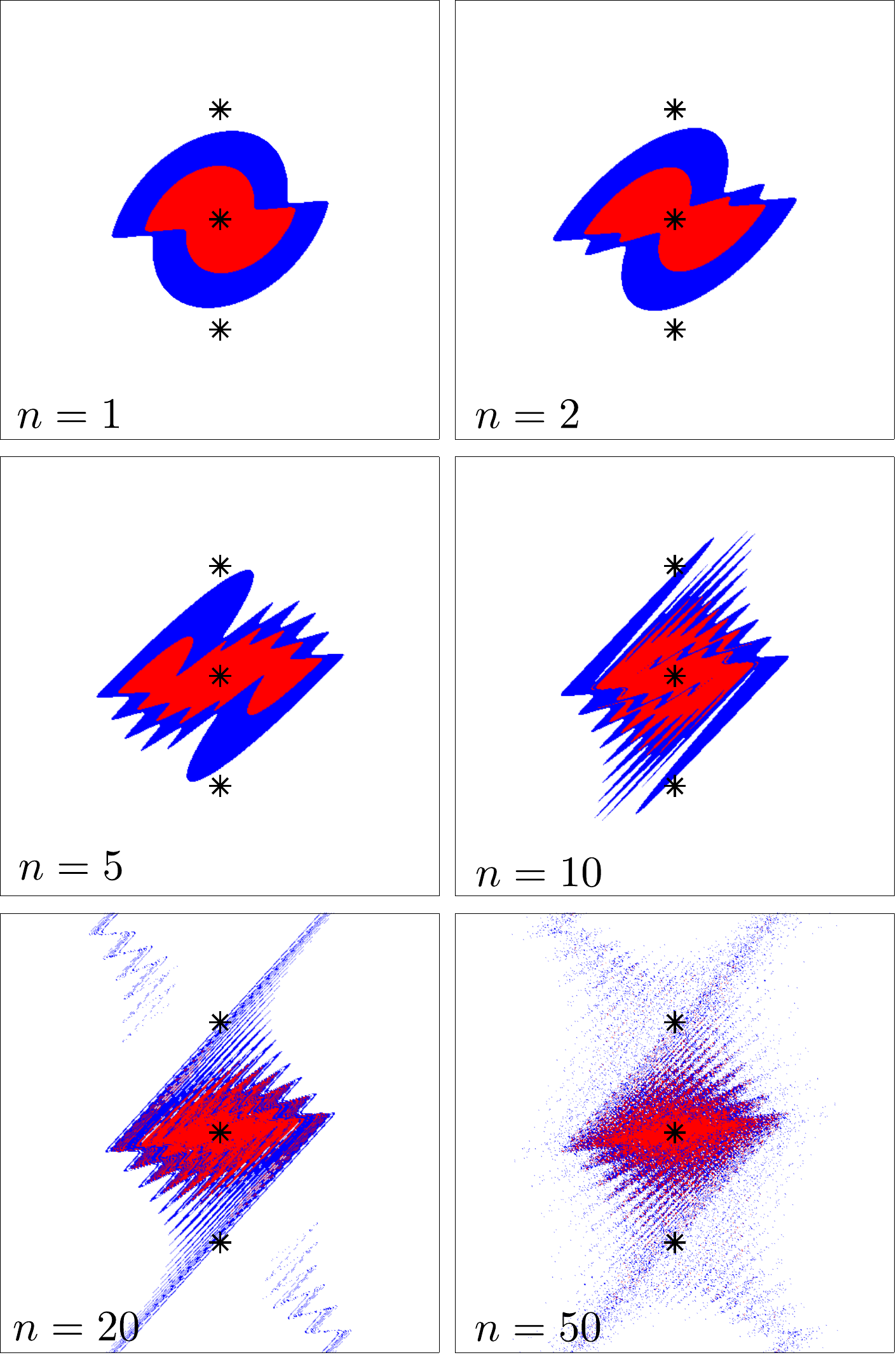}
  \end{minipage}\hfill
  \begin{minipage}[c]{0.35\textwidth}
    \caption{
       The dye trace simulation corresponding to Fig.~\ref{fig:scs_dye_trace} except using the smoothed map $\Lambda_b$ with $b=50$. Rather than stretching and CS, SF is the mixing mechanism. The period-1 points are shown as black stars, at $(0,\pm 1)$ they are hyperbolic, and the point at $(0,0)$ is elliptic. (Multimedia view).
    } \label{fig:smooth_dye_trace}
  \end{minipage}
\end{figure*}


The Lagrangian discontinuity in the CSS map is caused by the cutting map $C$. By approximating the cut with a smooth sequence of progressively sharper deformations we can analyse the transition of smooth systems with mixing controlled by SF towards a system with Lagrangian discontinuities and CS. We replace the map $C$ in the CSS map with the smoothed map
\begin{align}
&\mathcal{C}_b(x,y) = (x+ a \, g_b(y),y), \, \text{where} \\
&g_b(y) = \begin{cases}
A \tanh \left(b \left(y+2 \left| \frac{a}{\gamma _1}\right| \right)\right)-2  & , \,y < -\left| \frac{a}{\gamma _1}\right| \\
  A \tanh (b y)  & ,\,|y| \leq \left| \frac{a}{\gamma _1}\right| \\
  A \tanh \left(b \left(y-2 \left| \frac{a}{\gamma _1}\right| \right)\right)+2  &, \, y > \left| \frac{a}{\gamma _1}\right|
\end{cases} \\
& A = \coth \left(b \left| \frac{a}{\gamma _1}\right| \right).
\end{align}
As shown in Fig.~\ref{fig:smooth_cut_fn}, the $C^1$-continuous functions $g_b(y)$ converge pointwise to $\text{sgn}(y)$ as $b \to \infty$. Therefore the maps $\mathcal{C}_b$ converge pointwise to the cutting map $C$, and the associated smoothed CSS map $\Lambda_b = S_v S_h \mathcal{C}_b$ only has Lagrangian discontinuities in the limit $b\to \infty$. 

Comparing the dye trace simulations in Figure~\ref{fig:smooth_dye_trace_origin} for a fixed number of iterations $n=6$, the only difference between the CSS map ($b\to \infty$) and the smoothed map $\Lambda_b$ with large but finite $b$ is that material that is cut by the CSS map remains connected by thin striations. The striations occur along lines that coincide with the images of the Lagrangian discontinuity (Figure~\ref{fig:smooth_dye_trace_origin}b) and become infinitely thin as $b\to \infty$. This is expected since this web consists of the points that experience discontinuous deformation. Therefore transport structures for the smooth and discontinuous maps will converge everywhere except the web of post-images of the Lagrangian discontinuity, $D^u$, a set of measure zero.

In order to analyse the transition as $b \to \infty$ we consider the Poincar\'{e} sections and period-1 points. Fig.~\ref{fig:smooth_cut} demonstrates the convergence of Poincar\'{e} sections for the smooth systems towards the discontinuous limits Fig.~\ref{fig:scs_no_rotation}g,h. Unlike the CSS map, in each case at low values of $b$ the chains of periodic points alternate between elliptic (triangles) and hyperbolic (circles), as anticipated by the Poincar\'{e}--Birkhoff theorem. Focusing on the period-1 points for $\gamma_2 = -0.2$ (top row of Fig.~\ref{fig:smooth_cut}), as for the CSS map there are the two elliptic points at $(0, \pm 1)$. In contrast to the CSS map that has pseudo-hyperbolic points at the origin and on the periodic boundary at $(0,-|a/\gamma_1|)$, the smoothed maps $\Lambda_b$ have genuine hyperbolic points. Hence mixing in the chaotic sea between KAM islands is generated by SF rather than CS. As $b\to \infty$, the magnitude of the shear about the origin increases, and the eigenvalues $\lambda_1=1/\lambda_2$ approach $\infty$ and $0$ (Appendix~\ref{app:ppoints_smooth}), meaning infinite expansion and contraction. The corresponding eigenvectors give the directions of expansion and contraction, in the limit $b\to\infty$ they approach $(1,\gamma_2)$ and $(1,0)$ which correspond to the first image and preimage in the webs of Lagrangian discontinuities, and the only two lines that are tangent to both of the period-1 elliptic islands for the CSS map. 

Conversely, when the vertical shear is reversed (bottom row of Fig.~\ref{fig:smooth_cut}) there are two period-1 hyperbolic points at $(0,\pm 1)$ as for the CSS map, but there are two more period-1 points, at the origin and on the periodic boundary at $(0,-|a/\gamma_1|)$. These two new period-1 points share the same characteristics for each set of flow parameters. For low values of the smoothing parameter $b$, they are elliptic and the corresponding KAM islands are bounded by the parallel heteroclinic connections of unstable manifolds associated with the hyperbolic points. As $b$ increases, the heteroclinic connections become transverse, creating a chaotic sea around the KAM islands, gradually engulfing them. The KAM islands become increasingly thin, to the point where they undergo period-doubling bifurcations, each creating a period-1 hyperbolic point and two period-2 elliptic points. For the case shown with $(a,\gamma_1,\gamma_2)=(-0.2,0.2,0.2)$, this bifurcation occurs around $b\approx 101$ (Appendix~\ref{app:ppoints_smooth}). As in the $\gamma_2 = -0.2$ case, the eigenvalues approach $\infty$ and $0$, with eigenvectors converging to $(1,\gamma_2)$ and $(1,0)$, and associated stable and unstable manifolds converging to the webs of Lagrangian discontinuities $D^s,D^u$ respectively (Figure~\ref{fig:smooth_dye_trace_origin}d,e). These webs can therefore be considered as the analogues of the stable and unstable manifolds for systems with discontinuous deformations. 

A fundamental difference between the stable and unstable manifolds (Fig.~\ref{fig:smooth_manifolds}) for the CSS map compared to the smoothed map $\Lambda_b$ is that the Lagrangian discontinuity cuts both manifolds into disconnected pieces. Material lines such as the stable and unstable manifolds form impenetrable barriers to transport, but cutting them creates `gaps', allowing material to pass through freely. For sufficiently large $b$, almost all individual particles follow the same trajectories using the smoothed and discontinuous maps, however finite area material parcels have greater freedom in the presence of Lagrangian discontinuities. In the limit $b\to \infty$ the material barriers disappear, and material parcels are able to cross the destroyed barriers, whereas for finite $b$ high levels of stretching are required to ensure that barriers are not crossed. 

To consider the impact of smoothing the Lagrangian discontinuity on mixing characteristics we compare the dye trace simulation Fig.~\ref{fig:scs_dye_trace}a for the CSS map with its counterpart Fig.~\ref{fig:smooth_dye_trace} for the smoothed map $\Lambda_b$ with $b=50$. Both maps exhibit similar gross behaviour, however a key difference is that folds replace the pronounced cuts. The folds accumulate along the stable manifolds of the hyperbolic points (Fig.~\ref{fig:smooth_manifolds}) leading to strong SF. As $b$ increases the thickness of the folds decreases, so that in the limit $b\to \infty$ the folds have zero thickness, forming the lines of discontinuous deformation.

Considering a smoothed cut illustrates how non-Hamiltonian structures such as pseudo-elliptic points, pseudo-hyperbolic points and the webs of Lagrangian discontinuities are analogous to the structures found in Hamiltonian systems. Systems with highly localised shears share many of the same properties as systems with Lagrangian discontinuities, but there is a set of measure zero where transport structures will never agree. 

We have shown that in the RPM flow the Lagrangian discontinuities are a result of opening and closing valves combined with a slip boundary condition. In practice, fluid flows have no-slip boundaries, so fluid would always remain connected across valves. Instead of being cut, fluid will experience highly localised shear depending on the thickness of the boundary layer. We therefore expect that with no-slip boundary conditions in the RPM flow, the structures seen in Fig.~\ref{fig:RPM_p-sections}c would be replaced by their smoothed analogue as in Fig.~\ref{fig:smooth_cut}, where the thickness of the boundary layer acts as the smoothing parameter $b$.

\section{Conclusions}

Cutting by Lagrangian discontinuities drastically alters how mixing and transport can arise in materials and systems that have slip planes, shear bands or valves and wall slip. Cutting opens up a wider range of possibilities for the long-time organization of material by increasing the topological freedom of Lagrangian transport. Using a simple map consisting of cutting and shearing motions we have found a novel mixing mechanism that arises in general systems that combine stretching and CS, which we call a pseudo-elliptic island. Within these islands strong mixing occurs, an impossibility for CS-only systems and SF systems without folding. The map can also produce fractal tilings of classical non-mixing islands with a greater density of islands than is possible in a classical SF system, resulting in slower transport and weak mixing. We have only considered a few of the possibilities for maps composed of simple shear, cut and rotation deformations. Further study of the full array of possibilities could reveal fundamentally different structures that may appear in physical systems.

Our map demonstrates qualitatively similar mixing and transport as that seen in a realistic example flow for relevant ranges of the control parameters. We have shown that even with a smooth incompressible base flow, the opening and closing of valves combined with a slip boundary condition creates Lagrangian discontinuities, leading to transport behaviour that cannot be found in classical Hamiltonian systems. Slip walls are necessary if the opening and closing of valves occurs on the boundary, but Lagrangian discontinuities will also arise in flows with no-slip boundary conditions if the fluid is injected into and extracted from points in the domain away from boundaries.

The standard methods used for studying SF systems -- periodic point analysis, dye trace simulations -- are still relevant in systems with Lagrangian discontinuities, but new methods are needed to fully understand them. We have introduced the webs of Lagrangian discontinuities as a more comprehensive tool. These webs determine the location and stability of periodic points, the nature of the cutting mechanism, and the locations of new structures such as pseudo-periodic points.

While we have used 2D examples, extension of this analysis to fully 3D systems will be challenging. A third dimension creates even more topological possibilities for coherent structures, for instance the web of Lagrangian discontinuities is made from 1D curves in 2D systems but in 3D there will be a web of interlaced 2D surfaces. The introduction of a Lagrangian discontinuity is anticipated to create fundamentally different structures to those observed in 2D systems.

Future study should focus on developing a complete framework for 2D transport in the presence of Lagrangian discontinuities. Studying the full range of interactions between CS and SF.

\begin{acknowledgments}
L. Smith is funded by a Monash Graduate Scholarship and a CSIRO Top-up Scholarship.
\end{acknowledgments}

\appendix

\section{Particle Tracking in the RPM Flow} \label{app:particle_tracking}

Tracking individual particles within the RPM flow is essential for the production of Poincar\'{e} sections and dye trace simulations. Finding the path of a particle requires solving the advection equation~(\ref{eq:advection_eq}) either numerically or analytically. For the RPM flow the velocity field $\bm{v}$ can be found analytically \cite{Lester}, making computation significantly easier. While an analytic solution to equation~(\ref{eq:advection_eq}) is impossible for the RPM flow, a pseudo-analytic method is used to track particles in a similar manner to Lester et al\cite{Lester}. The main difference being the use of the potential function $\Phi$ and streamfunction $\Psi$ as orthogonal coordinates, instead of using the polar angle $\theta$ and $\Psi$ as coordinates which are parallel along the $y$-axis. Using the streamfunction as a coordinate is beneficial since it is a constant of motion, and the potential function is a canonical choice since it must be orthogonal to the streamfunction. Conversion between cartesian coordinates and $(\phi,\psi)$ coordinates is performed via polar coordinates
\begin{align}
\begin{split}
&(x,y) \leftrightarrow (r,\theta) \leftrightarrow (\phi,\psi) \\
&r(\phi,\psi) = \sqrt{1-\frac{2 \cos (\psi )}{\cos (\psi )+\cosh (\phi )}} \\
&\theta(\phi,\psi) = -\frac{\left| \phi \right|}{\phi }   \cos ^{-1}\left(\frac{\sin (\psi
   )}{\sqrt{\cosh ^2(\phi )-\cos ^2(\psi )}}\right) \\
&\phi(r,\theta) = \frac{1}{2} \log \left(\frac{r^2-2 r \sin (\theta )+1}{r^2+2 r \sin
   (\theta )+1}\right) \\
&\psi(r,\theta) = \tan ^{-1}\left(\frac{2 r \cos (\theta )}{1-r^2}\right).
\label{eq:coord_conv} 
\end{split}
\end{align}
 The advection equation in these new coordinates is given by
\begin{equation}
\frac{d\Phi}{dt} = \big( \cos(\Psi) + \cosh(\Phi) \big)^2, \quad \frac{d\Psi}{dt} = 0.
\label{eq:advection_eq_orth}
\end{equation}
We have therefore reduced the system to one dimension, though this differential equation is still insoluble. On the other, hand the equation 
\begin{equation}
\frac{dt}{d\Phi}=\frac{1}{d\Phi/dt}
\label{eq:inverse_advection}
\end{equation} 
has the analytic solution
\begin{align}
\begin{split}
&t_{\text{adv}}(\phi,\psi) = \csc ^2\psi  \Big[-2 \cot \psi  \tan ^{-1}\left(\tan
   \frac{\psi }{2} \tanh \frac{\phi
   }{2}\right) \\
   &\!\!+\frac{\sin \frac{\psi }{2} + \sinh \frac{\phi }{2}}{\sin \psi}
   \left(\text{sech}\left(\frac{\phi -i \psi
   }{2} \right)+\text{sech}\left(\frac{\phi +i \psi
   }{2} \right)\right)\Big]
\label{eq:tadv}
\end{split}
\end{align}
which gives the advection time from the $x$-axis ($\Phi = 0$) to the point $(\phi,\psi)$ along the streamline $\Psi=\psi$. The residence time of each streamline can then be computed as 
\begin{align}
\begin{split}
t_{\text{res}}(\psi) &= 2 \lim_{\phi \to \infty} t_{\text{adv}}(\phi,\psi) \\
&= -2 (\psi  \cot (\psi )-1) \csc ^2(\psi ).
\label{eq:tres}
\end{split}
\end{align}
The advection of a particle for the time period $T$ can be expressed as a map $(\phi,\psi)\to (\phi',\psi)$, where the new value of the potential function satisfies
\begin{equation}
t_{\text{adv}}(\phi',\psi) + \frac{t_{\text{res}}(\psi)}{2} =  t_{\text{adv}}(\phi,\psi) + \frac{t_{\text{res}}(\psi)}{2} + T\!\!\!\!\! \mod\! t_{\text{res}}(\psi).
\label{eq:t_map}
\end{equation}  
However, the function $t_{\text{adv}}$ is not invertible, so there is no analytic solution for $\phi'$. We use Newton's root finding method to solve equation~(\ref{eq:t_map}) to machine precision accuracy. The new coordinates are then converted to cartesian coordinates via equation~(\ref{eq:coord_conv}).

\section{Periodic Point Analysis}

Periodic points and their stability play a pivotal role in the organisation of particle transport, and lower order (smaller period) points play a greater role than higher order points. We therefore focus on finding the period-1 points of maps $f(x)$, which satisfy $f(x)=x$. The local stability of period-1 points is determined by the eigenvalues of the deformation tensor $\bm{F} = (\partial f_i/\partial x_j)$. For area-preserving maps the eigenvalues must satisfy $\lambda_1 \lambda_2 = 1$. If the eigenvalues are real, then $\lambda_1 = 1/ \lambda_2$, leading to a direction of contraction and a direction of expansion. In this case the periodic point is called hyperbolic. The only other possibility for an area-preserving map is that the eigenvalues form a complex conjugate pair $\lambda_{1,2} = \cos\theta \pm i \sin \theta$, in which case there is a rotation about the periodic point, and the point is called elliptic. 

\subsection{The CSS map}

To find and classify the period-1 points of the CSS map we write the map explicitly as
\begin{align}
\begin{split}
&\Lambda(x,y) = (x',y')  \\
&x' = x + \gamma_1 y + a \cdot \text{sgn}(y) \\
&y' = \left( \gamma_2 x' + y + \frac{2}{|\gamma_1'|} \!\!\!\! \mod \frac{4}{|\gamma_1'|} \right) - \frac{2}{|\gamma_1'|} ,
\label{eq:CSS_explicit}
\end{split}
\end{align}
where $\gamma_i'=\gamma_i/a$. This includes the periodic boundaries at $y=\pm 2/\gamma_1'$. Period-1 points must therefore satisfy the pair of equations
\begin{align}
\begin{split}
&x + \gamma_1 y + a \cdot \text{sgn}(y) = x, \\
&\left(\gamma_2 x + y + \frac{2}{|\gamma_1'|} \!\!\!\! \mod \frac{4}{|\gamma_1'|} \right) - \frac{2}{|\gamma_1'|} = y.
\end{split}
\end{align}
The second equation simplifies to
\begin{equation}
\gamma_2 x + y + \frac{2}{|\gamma_1'|} = y + \frac{2}{|\gamma_1'|} + n \frac{4}{|\gamma_1'|}, \quad n \in \mathbb{Z}
\end{equation}
which implies that 
\begin{equation}
x=n\frac{4}{|\gamma_1'|\gamma_2}, \quad |y|= -\frac{a}{\gamma_1}, \quad n \in \mathbb{Z}.
\end{equation}
Therefore period-1 points only exist when $\gamma_1'=\gamma_1/a<0$, and in these cases occur at the points $(n\frac{4}{|\gamma_1'|\gamma_2},\pm 1/\gamma_1')$.

For the CSS map the deformation tensor is
\begin{equation} \label{eq:css_def_tensor}
\bm{F} = \left( \begin{matrix}
1 &\gamma_1 \\
\gamma_2 &1+\gamma_1\gamma_2
\end{matrix} \right), \quad y\neq 0, \pm \frac{2}{\gamma_1'}
\end{equation}
and is undefined if $y=0,\pm 2/\gamma_1'$ due to discontinuities created by the cutting transformation and periodic boundary. The eigenvalues of $\bm{F}$ are
\begin{equation} \label{eq:def_tensor_eigen}
\lambda_{1,2} = \frac{1}{2}\left( 2 + \gamma_1\gamma_2 \pm \sqrt{\gamma_1\gamma_2(4+\gamma_1\gamma_2)} \right).
\end{equation}
Assuming that $\gamma_1>0$ and $\gamma_2 > -4/\gamma_1$, the eigenvalues are real when $\gamma_2>0$ and form a complex conjugate pair when $\gamma_2<0$. In terms of stability this means that the period-1 points are hyperbolic when $\gamma_2>0$ and they are elliptic when $\gamma_2<0$.

\subsection{The non-linear CSS map}

The function $f(x)$ in the quadratic vertical shear $S_{nl}$ can be written as $f(x) = \alpha x^2 - \beta$. The period-1 points of the non-linear CSS map can be found in a similar manner to those in the linear CSS map. The map can be written explicitly as
\begin{align}
\begin{split}
&\Lambda_2(x,y) = (x',y')  \\
&x' = x + \gamma_1 y + a \cdot \text{sgn}(y) \\
&y' = \left( \alpha (x')^2 - \beta + y + \frac{2}{|\gamma_1'|} \mod \frac{4}{|\gamma_1'|} \right) - \frac{2}{|\gamma_1'|}.
\label{eq:CSS2_explicit}
\end{split}
\end{align}
Period-1 points satisfy $(x',y')=(x,y)$, which is true when
\begin{equation}
x=\pm \sqrt{\frac{1}{\alpha}\left( \beta + n \frac{4}{|\gamma_1'|} \right) }, \quad |y|= -\frac{a}{\gamma_1} , \quad n \in \mathbb{Z}.
\end{equation}
This can only occur when $\frac{1}{\alpha}\left( \beta + n \frac{4}{|\gamma_1'|} \right)\geq 0$ and $\gamma_1'<0$.
The deformation tensor $\bm{F}_2$ can be computed at every point except along the lines $y=0,\pm 2/\gamma_1'$:
\begin{equation}
\bm{F}_2(x,y) = \left( \begin{matrix}
1 &\gamma_1 \\
2\alpha x' &1+2\gamma_1\alpha x'
\end{matrix} \right).
\end{equation}
It can be seen that $\bm{F}_2$ is equivalent to $\bm{F}$ but with $\gamma_2$ replaced with $2\alpha x'$. Assuming that $\gamma_1>0$ and $\alpha>0$, we can therefore say that the period-1 points are elliptic when $x<0$ and hyperbolic when $x>0$. 

\subsection{Smoothed CSS map} \label{app:ppoints_smooth}

Essentially the same analysis as for the CSS map applies to the smoothed map at the points $(0,\pm 1)$. The value of the smoothing parameter $b$ has some effect on the values of the eigenvalues and eigenvectors of the deformation tensor, but the nature of the points remains the same.

For the period-1 points at the origin and on the periodic boundary at $(0,-2|a/\gamma_1|)$, direct computation of the deformation tensor $\bm{F}_b=(\partial \Lambda_b^i/\partial x_j)$ yields
\begin{equation} \label{eq:smooth_def_tensor}
\bm{F}_b (x,y) = \left( \begin{matrix}
1 &\beta \\
\gamma_2 &1+\beta\gamma_2
\end{matrix} \right)
\end{equation}
where
\begin{equation}
\beta =  a b \coth \left(b \left| \frac{a}{\gamma _1}\right| \right)+\gamma _1
\end{equation} 
is the net horizontal shear at the origin from the maps $\mathcal{C}_b$ and $S_h$. This deformation tensor is in the same form as equation~\ref{eq:css_def_tensor}, and thus the two eigenvalues are
\begin{equation}
\lambda_{1,2} = \frac{1}{2} \left( 2+ \beta \gamma_2 \pm  \sqrt{\beta \gamma_2  (4 + \beta \gamma_2 ) }\right).
\end{equation}
As $b\to \infty$, the eigenvalues converge to $0$ and $-\infty$ respectively.

The nature of the periodic point is determined by the discriminant
\begin{equation}
\Delta = \beta \gamma_2 (4+ \beta\gamma_2),
\end{equation} 
for $\Delta > 0$ the period-1 point is hyperbolic, for $\Delta <0$ it is elliptic and when $\Delta = 0$ it is degenerate. Assuming that $\gamma_1>0$, $\gamma_2\neq 0$ and $a<0$, there are a number of cases. The discriminant will be zero when
\begin{equation} \label{eq:smooth_disc_sol}
\beta = 0, \quad \text{or} \quad \beta = \frac{-4}{\gamma_2}.
\end{equation} 
Restricting to the case when $\gamma_1=-a$, then $\beta=0$ is equivalent to $b\coth b = 1$, implying that $b=0$. Therefore the period-1 point becomes degenerate as the smoothing parameter $b \to 0$. This can also be seen by the fact that as $b\to 0$, $g_b(y) \to y$ so $\mathcal{C}_b \to S_h^{-1}$ and hence $\Lambda_b \to S_v$.

The other case in equation~(\ref{eq:smooth_disc_sol}), $\beta = \frac{-4}{\gamma_2}$, is equivalent to
\begin{equation} \label{eq:bif_sol}
b \coth \left(b \left|\frac{a}{\gamma_1}\right|\right) = \frac{-1}{a} \left( \frac{4}{\gamma_2} + \gamma_1 \right)
\end{equation}
which will have a solution provided the right hand side is greater than $1$. If a solution exists then the period-1 point will experience either a period-doubling or period-halving bifurcation. In the main text the cases used are $(a,\gamma_1,\gamma_2)=(-0.2,0.2,\pm 0.2)$. For $\gamma_2=0.2$, the right hand side of equation~\ref{eq:bif_sol} is equal to $101$, and hence there is a period-doubling bifurcation of the elliptic point at $b\approx 101$. On the other hand, for $\gamma_2=-0.2$, the right hand side of equation~\ref{eq:bif_sol} is equal to $-99$ and hence there is no solution. In this case the point remains hyperbolic for all values of $b$.

The eigenvectors of the deformation tensor $\bm{F}_b$ corresponding to the eigenvalues $\lambda_{1,2}$ are given by
\begin{equation}
\bm{v}_1 = \left( 1, \frac{\gamma_2}{1-\lambda_2} \right), \, \text{and } 
\bm{v}_2 = \left( 1, \frac{\gamma_2}{1-\lambda_1} \right).
\end{equation} 
Hence as $b\to \infty$ the eigenvectors converge to $(1,0)$ and $(1,\gamma_2)$ respectively.

\end{document}